\documentclass[twocolumn,superscriptaddress,pre]{revtex4-1}

\newcommand{\tw}{\ensuremath{t_\mathrm{w}}\xspace}
\newcommand{\Tc}{\ensuremath{T_\mathrm{c}}\xspace}
\newcommand{\ee}{\text{e}}
\usepackage[T1]{fontenc}
\usepackage[utf8]{inputenc}
\usepackage{graphicx}
\usepackage{amsmath}
\usepackage{amssymb}
\usepackage{color}
\usepackage{xspace}
\usepackage{multirow}
\usepackage{hyperref}
\makeatletter
\newcommand*{\balancecolsandclearpage}{%
  \close@column@grid
  \cleardoublepage
  \twocolumngrid
}
\makeatother
\graphicspath{{./figures/}{./}}

\begin{document}

\title{Aging rate of spin glasses from simulations matches
experiments}

\author{M.~Baity-Jesi}\affiliation{Department of Chemistry, Columbia University, New York, NY 10027, USA}\affiliation{Instituto de Biocomputaci\'on y F\'{\i}sica de Sistemas Complejos (BIFI), 50018 Zaragoza, Spain}

\author{E.~Calore}\affiliation{Dipartimento di Fisica e Scienze della Terra, Universit\`a di Ferrara e INFN, Sezione di Ferrara, I-44122  Ferrara, Italy}

\author{A.~Cruz}\affiliation{Departamento  de F\'\i{}sica Te\'orica, Universidad de Zaragoza, 50009 Zaragoza, Spain}\affiliation{Instituto de Biocomputaci\'on y F\'{\i}sica de Sistemas Complejos (BIFI), 50018 Zaragoza, Spain}

\author{L.A.~Fernandez}\affiliation{Departamento  de F\'\i{}sica Te\'orica, Universidad Complutense, 28040 Madrid, Spain}\affiliation{Instituto de Biocomputaci\'on y F\'{\i}sica de Sistemas Complejos (BIFI), 50018 Zaragoza, Spain}

\author{J.M.~Gil-Narvion}\affiliation{Instituto de Biocomputaci\'on y F\'{\i}sica de Sistemas Complejos (BIFI), 50018 Zaragoza, Spain}

\author{A.~Gordillo-Guerrero}\affiliation{Departamento de  Ingenier\'{\i}a El\'ectrica, Electr\'onica y Autom\'atica, U. de Extremadura, 10003, C\'aceres, Spain}\affiliation{Instituto de Computaci\'on Cient\'{\i}fica Avanzada (ICCAEx), Universidad de Extremadura, 06006 Badajoz, Spain}\affiliation{Instituto de Biocomputaci\'on y F\'{\i}sica de Sistemas Complejos (BIFI), 50018 Zaragoza, Spain}

\author{D.~I\~niguez}\affiliation{Instituto de Biocomputaci\'on y F\'{\i}sica de Sistemas Complejos (BIFI), 50018 Zaragoza, Spain}\affiliation{Fundaci\'on ARAID, Diputaci\'on General de Arag\'on, Zaragoza, Spain}

\author{A.~Maiorano}\affiliation{Dipartimento di Fisica, Sapienza
  Universit\`a di Roma, 
  I-00185 Rome, Italy}\affiliation{Instituto de Biocomputaci\'on y F\'{\i}sica de Sistemas Complejos (BIFI), 50018 Zaragoza, Spain}

\author{E.~Marinari}\affiliation{Dipartimento di Fisica, Sapienza
  Universit\`a di Roma, INFN, Sezione di Roma 1, and CNR-Nanotec,
  I-00185 Rome, Italy}

\author{V.~Martin-Mayor}\affiliation{Departamento  de F\'\i{}sica Te\'orica, Universidad Complutense, 28040 Madrid, Spain}\affiliation{Instituto de Biocomputaci\'on y F\'{\i}sica de Sistemas Complejos (BIFI), 50018 Zaragoza, Spain}

\author{J.~Moreno-Gordo}\affiliation{Instituto de Biocomputaci\'on y F\'{\i}sica de Sistemas Complejos (BIFI), 50018 Zaragoza, Spain}\affiliation{Departamento  de F\'\i{}sica Te\'orica, Universidad de Zaragoza, 50009 Zaragoza, Spain}

\author{A.~Mu\~noz-Sudupe}\affiliation{Departamento  de F\'\i{}sica Te\'orica, Universidad Complutense, 28040 Madrid, Spain}\affiliation{Instituto de Biocomputaci\'on y F\'{\i}sica de Sistemas Complejos (BIFI), 50018 Zaragoza, Spain}

\author{D.~Navarro}\affiliation{Departamento de Ingenier\'{\i}a, Electr\'onica y Comunicaciones and I3A, U. de Zaragoza, 50018 Zaragoza, Spain}

\author{G.~Parisi}\affiliation{Dipartimento di Fisica, Sapienza
  Universit\`a di Roma, INFN, Sezione di Roma 1, and CNR-Nanotec,
  I-00185 Rome, Italy}

\author{S.~Perez-Gaviro}\affiliation{Centro Universitario de la Defensa, Carretera de Huesca s/n, 50090 Zaragoza, Spain}\affiliation{Instituto de Biocomputaci\'on y F\'{\i}sica de Sistemas Complejos (BIFI), 50018 Zaragoza, Spain}\affiliation{Departamento  de F\'\i{}sica Te\'orica, Universidad de Zaragoza, 50009 Zaragoza, Spain}

\author{F.~Ricci-Tersenghi}\affiliation{Dipartimento di Fisica, Sapienza
  Universit\`a di Roma, INFN, Sezione di Roma 1, and CNR-Nanotec,
  I-00185 Rome, Italy}

\author{J.J.~Ruiz-Lorenzo}\affiliation{Departamento de F\'{\i}sica, Universidad de Extremadura, 06006 Badajoz, Spain}\affiliation{Instituto de Computaci\'on Cient\'{\i}fica Avanzada (ICCAEx), Universidad de Extremadura, 06006 Badajoz, Spain}\affiliation{Instituto de Biocomputaci\'on y F\'{\i}sica de Sistemas Complejos (BIFI), 50018 Zaragoza, Spain}

\author{S.F.~Schifano}\affiliation{Dipartimento di Matematica e Informatica, Universit\`a di Ferrara e INFN, Sezione di Ferrara, I-44122 Ferrara, Italy}

\author{B.~Seoane}\affiliation{Laboratoire de physique th\'eorique, D\'epartement de physique de l'ENS, \'Ecole normale sup\'erieure, PSL Research University, Sorbonne Universit\'e, CNRS, 75005 Paris, France}\affiliation{Instituto de Biocomputaci\'on y F\'{\i}sica de Sistemas Complejos (BIFI), 50018 Zaragoza, Spain}

\author{A.~Tarancon}\affiliation{Departamento  de F\'\i{}sica Te\'orica, Universidad de Zaragoza, 50009 Zaragoza, Spain}\affiliation{Instituto de Biocomputaci\'on y F\'{\i}sica de Sistemas Complejos (BIFI), 50018 Zaragoza, Spain}

\author{R.~Tripiccione}\affiliation{Dipartimento di Fisica e Scienze della Terra, Universit\`a di Ferrara e INFN, Sezione di Ferrara, I-44122 Ferrara, Italy}

\author{D.~Yllanes}\email{dyllanes@syr.edu}\affiliation{Department of Physics and Soft and Living Matter Program, Syracuse University, Syracuse, NY, 13244}\affiliation{Instituto de Biocomputaci\'on y F\'{\i}sica de Sistemas Complejos (BIFI), 50018 Zaragoza, Spain}

\collaboration{Janus Collaboration}

\date{\today}

\begin{abstract}
Experiments on spin glasses can now make precise measurements of the exponent
$z(T)$ governing the growth of glassy domains, while our computational
capabilities allow us to make quantitative predictions for experimental scales.
However, experimental and numerical values for $z(T)$ have differed. We use new
simulations on the Janus II computer to resolve this discrepancy, finding a
time-dependent $z(T, \tw)$, which leads to the experimental value through mild
extrapolations. Furthermore, theoretical insight is gained by studying a
crossover between the $T = \Tc$ and $T = 0$ fixed points.
\end{abstract}

\maketitle 

The study of spin glasses (SGs)~\cite{mydosh:93,young:98} has long been  a
key problem in statistical mechanics, providing ideas that have born fruit in
fields as diverse as econophysics, biology or optimization in computer science. 
From a fundamental point of view, SGs are paradigmatic 
as the most approachable model for glassy behavior, both 
experimentally and theoretically.  However, despite this relative simplicity, 
SG experiments and theory have traditionally developed separately,
for practical and conceptual reasons. On the one hand, numerical simulations
were not long enough to reach experimental times, while experiments
were not precise enough or even able to measure  key physical quantities.
On the other hand, experimental samples are perennially out of equilibrium,
while theory mostly focuses on the (unreachable) equilibrium phase.

In a typical experiment, the system is rapidly cooled to a
subcritical working temperature $T<\Tc$ and its off-equilibrium
evolution (aging) studied. As the waiting time \tw increases, the size
of the glassy domains is seen to grow as $\xi(\tw)\propto
\tw^{1/z(T)}$, with an exponent that is expected to behave as
$z(T)\simeq z(T_\text{c}) \Tc/T$~\cite{marinari:00}.  In traditional
experiments~\cite{joh:99}, based on the shift of the peak in the
relaxation rate $S(\tw)$, $z(T)$ was difficult to
measure. Fortunately, the availability of excellent samples
with a film geometry has suggested a new approach to the precision
measurement of $z_\text{c}=z(T)T/\Tc$~\cite{zhai:17}. The time that
$\xi(\tw)$ needs to saturate to the film thickness relates to the
activation energies~$\varDelta_\text{max}$~\cite{guchhait:14,guchhait:17}.
Varying the film thickness from 9 to 20~nm resulted in the measurement
$z_\text{c}\approx 9.62$~\cite{zhai:17}, very far from
the value predicted by numerical simulations $z_\text{c}=6.86(16)$~\cite{janus:08b},
$z_\text{c} = 6.80(15)$~\cite{lulli:15}.

Fortunately, recent theoretical progress makes it feasible to address
the above-mentioned disagreement.  A key development has been the
introduction of the Janus~\cite{janus:09,janus:12b} and
Janus~II~\cite{janus:14} computers, which have extended the numerical
exploration of the dynamics almost to the experimental
scale~\cite{janus:08b,janus:16}.  In addition, the introduction of
quantitative statics-dynamics dictionaries (first based on microscopic
quantities~\cite{janus:08b,janus:10,janus:10b} and more recently on
experimentally measurable features~\cite{janus:16}) has clarified the
relevance of the equilibrium phase for the off-equilibrium dynamics
and showed how to extrapolate simulations to the experimental scale.
Finally, the (macroscopic) experimental measurement of the size of
glassy domains was shown to be consistent with the (microscopic)
definition based on correlation functions~\cite{janus:17b}.

Here we resolve the discrepancy in $z_\text{c}$ by finding a (very mild) scale
dependence in the dynamical exponent $z\bigl(T,\xi(\tw)\bigr)$. We first recognize
that \emph{time} should be traded by \emph{length} scales.  Gentle
extrapolations to the relevant experimental scales of 20
nm~\cite{zhai:17} then reconcile the numerical and experimental
measurements.  Such a computation has been possible only thanks to new
data with unprecedented precision, achieved by reducing the
uncertainty due to thermal fluctuations, an issue that was typically
neglected in previous numerical work. From the theoretical point of
view, our study is based on a characterization of the crossover
between critical and low-temperature behavior. This is a very
important point, since it resolves a theoretical controversy on how
low a temperature must be studied to be free of critical effects, with
some authors choosing to work at very low $T$ at the expense of the
system sizes that it is possible to equilibrate (e.g.,~\cite{wang:17})
and others trying to find a tradeoff between temperature and system
size (e.g.,~\cite{janus:10}).

We consider the standard Edwards-Anderson
model~\cite{edwards:75}, defined on a three-dimensional cubic 
lattice of side $L=160$, on whose nodes we place spins $S_{\boldsymbol x} = \pm1$
that interact with their lattice nearest neighbors through
\begin{equation}\label{eq:H}
\mathcal H = - \sum_{\langle \boldsymbol x,\boldsymbol y\rangle} J_{\boldsymbol x \boldsymbol y} S_{\boldsymbol x} S_{\boldsymbol y}\, .
\end{equation}
For each disorder realization $\{J_{\boldsymbol x\boldsymbol y}\}$
(a sample), each of the quenched couplings $J_{\boldsymbol x\boldsymbol y}$
is $\pm1$ with $50\%$ probability. We shall refer to thin
CuMn films~\cite{zhai:17}, where the film thickness of 20~nm
translates to a distance of 38 lattice spacings (the
typical Mn-Mn distance is $5.3$\AA). 

Our systems are initialized with random orientations for the spins 
(representing a very high starting temperature) and immediately
quenched to the working temperature $T<\Tc=1.102(3)$~\cite{janus:13}. 
We then follow the evolution with the waiting time \tw (measured 
in units of full lattice sweeps) at constant temperature. For each 
sample $\{J_{\boldsymbol x\boldsymbol y}\}$ we simulate $N_\text{R}$
real replicas, evolving with different thermal noise. We estimate
our statistical errors with a jackknife method~\cite{amit:05} 
(including fit parameters~\cite{yllanes:11}).

\begin{figure}[t]
\includegraphics[width=\linewidth]{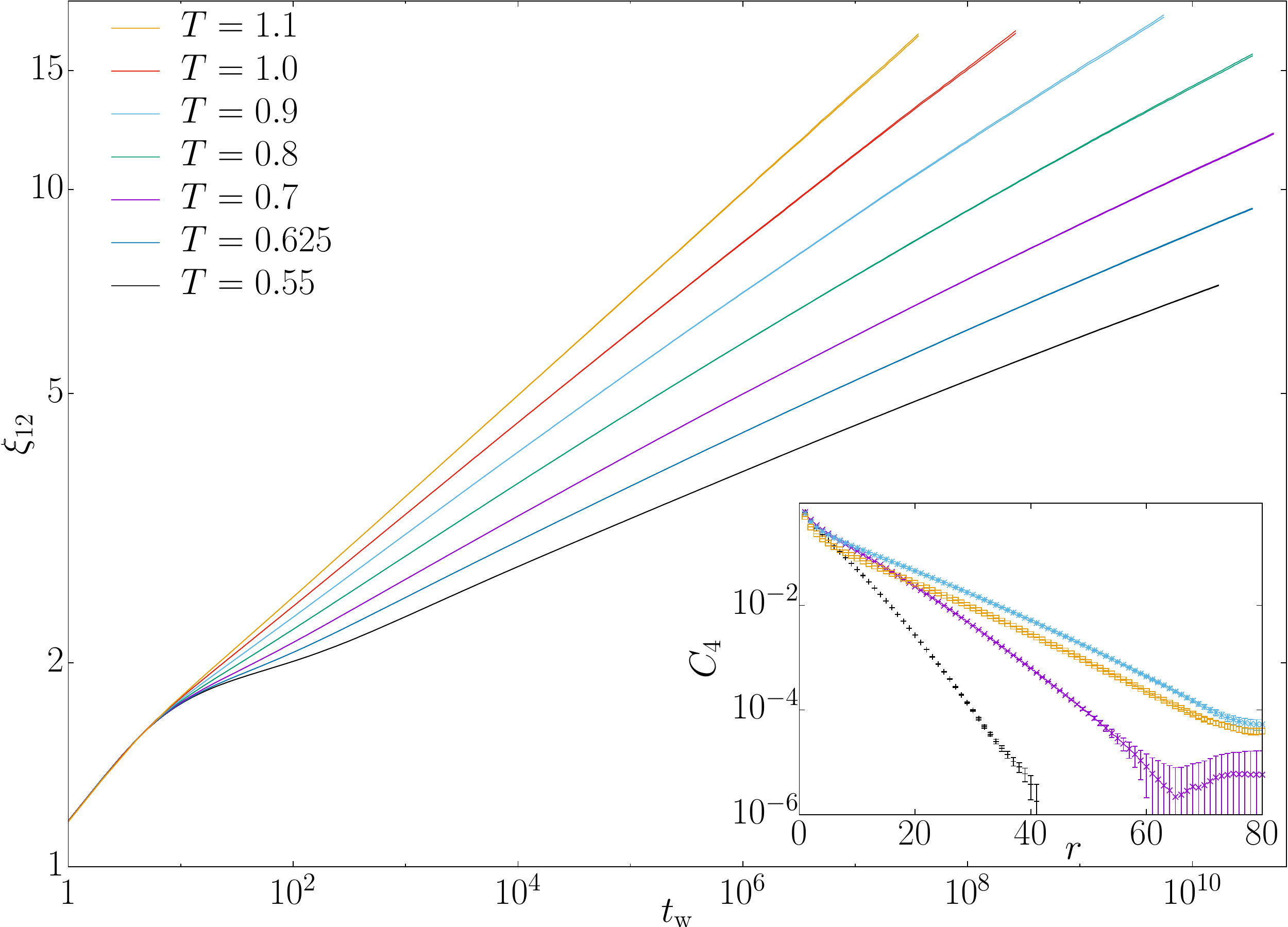}
\caption{Growth of the coherence length $\xi_{12}(T,\tw)$ with 
the waiting time \tw after a quench to temperature $T$ in a log-log 
scale [the critical temperature is $\Tc=1.102(3)$]. Given the smallness of the
statistical errors, instead of error bars we have plotted two lines for each
$T$, which enclose the error estimate.  At this scale, the curves seem linear
for long times, indicating a power-law growth but, see Fig.~\ref{fig:c2}, there
is actually a measurable curvature. \emph{Inset:} Spatial autocorrelation
function of the overlap field $C_4(T,r,\tw)$, plotted as a function of distance
at the last simulated time for several temperatures. Note the six orders of
magnitude in the vertical axis.
\label{fig:xi}}
\end{figure}

Our basic observable is the spatial autocorrelation 
of the overlap field (discussed in detail in~\cite{janus:09b}),
\begin{align}\label{eq:C4}
C_4(T,\boldsymbol r, \tw) &= \overline{\langle q^{(a,b)}(\boldsymbol x, \tw)
q^{(a,b)}(\boldsymbol x+\boldsymbol r , \tw)\rangle_T},\\
q^{(a,b)}(\boldsymbol x,\tw) &= S^{(a)}(\boldsymbol x,\tw)S^{(b)}(\boldsymbol x,\tw).
\end{align}
In these equations, the indices $(a,b)$ label the different 
real replicas; $\langle\cdots\rangle_{T}$ is the average over 
the thermal noise [in practice, an average over the
$(a,b)$ pairs] and $\overline{(\cdots)}$ is the average over the disorder.
In equilibrium simulations, by far the main source of error are the 
sample-to-sample fluctuations. Therefore, it has been customary 
to simulate the smallest $N_\text{R}$ that permits definitions 
such as~\eqref{eq:C4} and maximize the number $N_\text{S}$ of
samples. Instead, we have $N_\text{R}=256$ and $N_\text{S}=16$. This
choice, motivated to facilitate future studies of temperature 
chaos~\cite{janus:xx}, has proven crucial: contrary to our 
expectations, the increase in $N_\text{R}$ has produced 
a dramatic reduction of statistical errors (see Appendix~\ref{sec:Nr} for details).
As a result, we have been able to follow the decay of $C_4(T,r,\tw)$
over six decades (see inset to Fig.~\ref{fig:xi}).
A similar dramatic error reduction with high $N_\text{R}$
has also been seen in studies of the Gardner transition
in structural glasses~\cite{berthier:16,seoane:18}.

\newcommand{\dd}{\text{d}}
These correlation functions decay with distance as
\begin{equation}\label{eq:replicon}
 C_4(T,r,\tw) = r^{-\theta} f\bigl(r/\xi(T,\tw)\bigr), 
\end{equation}
so the growing $\xi$ can be computed through integral estimators~\cite{janus:08b,janus:09b}:
$I_k(T,\tw) = \int_0^\infty \dd r\ r^k C_4(T,r,\tw)$.
Then $\xi_{k,k+1}(T,\tw) = I_{k+1}(T,\tw)/I_k(T,\tw)$. As in recent
work~\cite{janus:09b,janus:17b,janus:16,fernandez:15} we use $k=1$
(see~\cite{fernandez:18a} for technical details). The resulting $\xi_{12}$ is
plotted in Fig.~\ref{fig:xi} for all our working temperatures.

\begin{figure}[t]
\includegraphics[width=\linewidth]{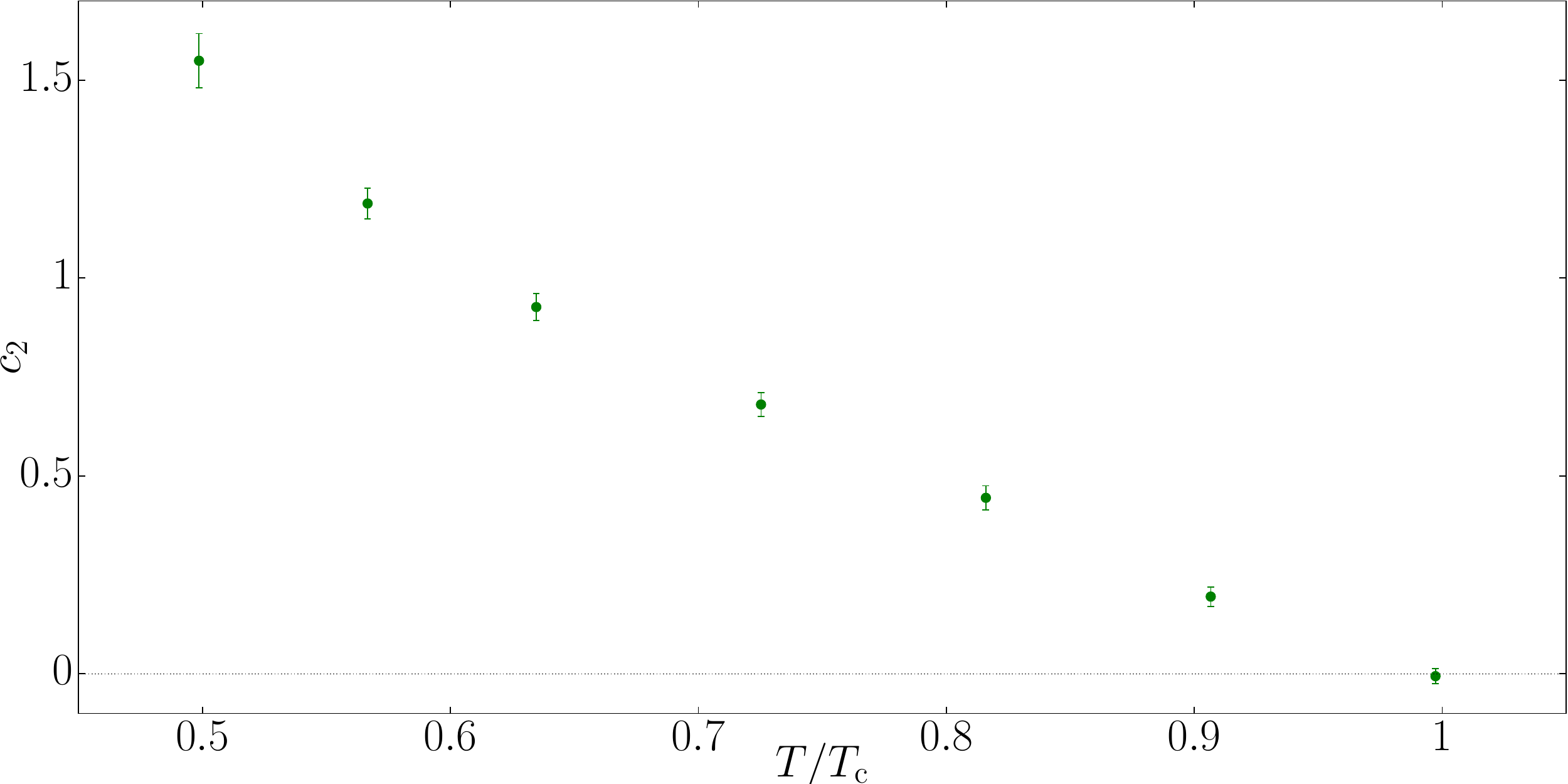}
\caption{Deviation of $\xi_{12}(\tw)$ from a simple power-law growth.
We plot the quadratic parameter $c_2$ in a fit to~\eqref{eq:c2}
(see Appendix~\ref{sec:parameters} for fitting parameters).
This quantity is zero at the critical point, but 
has a positive value at low temperatures, indicating that the growth 
of $\xi_{12}$ slows down over the simulated time range.
\label{fig:c2}}
\end{figure}

The numerical~\cite{janus:08b,janus:09b,fernandez:15} and
experimental~\cite{zhai:17} state of the art
describes the growth of $\xi_{12}$ with a power law,
\begin{equation}\label{eq:z}
\xi_{12}(T,\tw) \simeq A(T)\, \tw^{1/z(T)}\,.
\end{equation}
However, with our increased precision, \eqref{eq:z}~is no longer a
faithful representation of the dynamics. Indeed, if we switch to
$x=\log\xi_{12}$ as independent variable we can interpolate our data
as
\begin{equation}\label{eq:c2}
\log \tw(T,\xi_{12}) = c_0(T) + c_1(T)\ x + c_2(T)\ x^2.
\end{equation}
 Notice that $c_2=0$ would reduce to~\eqref{eq:z}, while $c_2>0$ would
indicate a slowing down of the dynamics for increasing
$\xi_{12}$. Indeed, see Fig.~\ref{fig:c2}, we find that $c_2$ vanishes
only at $T=\Tc$, with $z_\text{c} =z(T\!=\!\Tc) = 6.69(6)$~\footnote{
  Our result $z_\text{c} =z(T\!=\!\Tc) = 6.69(6)$ is significantly
  more accurate that $z_\text{c} = 6.80(15)$~\cite{lulli:15} and
  $z_\text{c}=6.86(16)$~\cite{janus:08b}, even though, unlike
  Ref.~\cite{janus:08b}, we allow for corrections to scaling,
  which increases the statistical error, see Appendix~\ref{sec:parameters}}. Of course,~\eqref{eq:c2}, useful as an interpolation, is not
suitable to extrapolate for longer times than simulated.  In order to
do that, we need some insight from theory~\footnote{A naive explanation for the
  curvature in $\xi_{12}(T,\tw)$ would be the existence of
  finite-size effects (see~\cite{janus:08b}). However,
  $c_2$ grows as we decrease $T$, while finite-size effects
  would be controlled by $\xi_{12}/L$, which is smaller for the lower
  temperatures. See Appendix~\ref{sec:finite_size_effects} for extensive checks 
  that our $L=160$ are safe.}.

We can gain much insight into the SG phase by considering the 
algebraic prefactor in~\eqref{eq:replicon}, determined by an exponent $\theta$.
At \Tc, $\theta=1+\eta$, where $\eta=-0.390(4)$~\cite{janus:13} is the
anomalous dimension. In the SG phase, there are differing expectations
for $\theta$ in the two main theoretical pictures. The droplet
description~\cite{mcmillan:83,bray:78,fisher:86}  expects coarsening domains
and therefore $\theta=0$. On the other hand, the replica symmetry breaking
(RSB) theory expects space-filling domains where $C_4$ vanishes at constant
$r/\xi_{12}$ as $\tw$ grows. In particular, $\theta$ is given by the replicon,
a critical mode analogous to magnons in Heisenberg ferromagnets
(see~\cite{janus:10b} for a detailed discussion). The
best previous numerical study of $\theta$~\cite{janus:09b}, found $\theta =
0.38(2)$, with a small $T$ dependence that was vaguely attributed to the effect of
the critical point.

\begin{figure}[t]
\includegraphics[width=\linewidth]{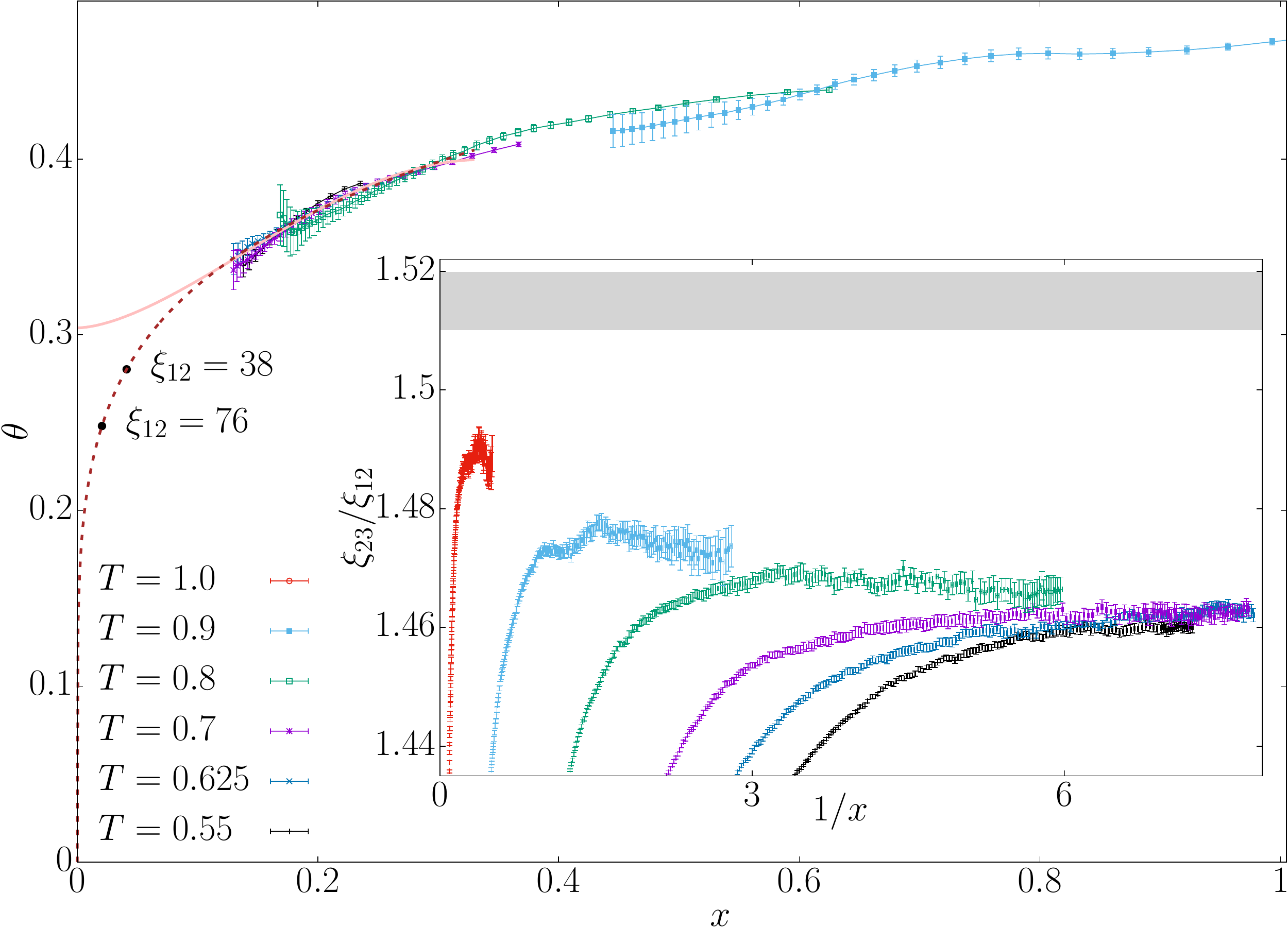}
\caption{Crossover between the $T=\Tc$ and the $T=0$ fixed points 
controlled by a Josephson length $\ell_\text{J}(T)$, with
$\ell_\text{J}(T)\propto(T_\text{c}-T)^{-\nu}$ close to \Tc (see text).
The relevant scaling variable is $x=\ell_\text{J}(T)/\xi_{12}$.
The \emph{inset} considers the ratio $\xi_{23}/\xi_{12}$ between two
definitions of the 
coherence length, which should be constant in the large-$\xi_{12}$ (or $x\to0$) limit.
For $T$ close to $\Tc$, this ratio initially grows, approaching the $T=\Tc$ value 
(represented by the thick gray line) and eventually relaxes towards the $T=0$
fixed point. 
\emph{Main plot:} Evolution of the replicon exponent $\theta$ 
for several temperatures. We
show two possible extrapolations to infinite $\xi_{12}$: one with finite
$\theta$, as expected in the RSB picture, and one with $\theta=0$, as expected
in the droplet picture. For the latter, we also show the extrapolated value for
the experimental scale corresponding to experiments 
in CuMn films~\cite{zhai:17}, which we estimate between 
$\xi_{12}=38$ and $\xi_{12}=76$.
\label{fig:replicon}}
\end{figure}

We can obtain $\theta$ by noticing that $I_2(T,\xi_{12}) \propto
\xi_{12}^{3-\theta}$. However, again we find that, while $\theta(\Tc)$
is compatible with $1+\eta$, for $T<\Tc$ we actually have
$\theta(T,\xi_{12})$, slowly decreasing as $\xi_{12}$ increases (or
$T$ decreases). This may seem an unsatisfactory result, since, in the
large-$\xi_{12}$ limit, $\theta(T,\xi_{12})$ should tend to a
$T$-independent constant (possibly zero). The simplest explanation is
that low values of $\xi_{12}$ are affected by the $T=\Tc$ fixed point
with $\theta \approx 0.61$ [an idea supported by the higher measured
$\theta(T,\xi_{12})$ for the higher $T$], while for
$\xi_{12}\to\infty$ we should see a crossover to the $T=0$ fixed
point, with an unknown $\theta(T=0)$ (see also~\cite{fernandez:15}).

In analogy with the ferromagnetic phase of the $O(N)$ model,
we can model this crossover in terms of a Josephson 
length~$\ell_\text{J}$~\cite{josephson:66}. Close to $\Tc$, 
this should grow as $\ell_\text{J}\propto(\Tc-T)^{-\nu}$, with
$\nu=2.56(4)$~\cite{janus:13}, while scaling corrections
are expected for the lowest temperatures~\footnote{In theory,
$\ell_\text{J} \propto [1+j_0(\Tc-T)^\nu+j_1(\Tc-T)^{\omega\nu}](\Tc-T)^{-\nu}$,
where we include analytic ($j_0$) and confluent ($j_1)$ scaling corrections 
with $\omega=1.12(10)$~\cite{janus:13}.
For Fig.~\ref{fig:replicon}, we have chosen $j_0$ and $j_1$ to obtain the best
collapse for the lowest temperatures.}.  If this
hypothesis is correct, our data for different temperatures should come together
when plotted in terms of a scaling variable $x=\ell_\text{J}/ \xi_{12}$.  We
test this scaling in the inset to Fig.~\ref{fig:replicon}, where we
consider the ratio $\xi_{23}/\xi_{12}$ between two different determinations of
the coherence length, which should be scale invariant in the large-$\xi_{12}$
limit (different definitions of $\xi$ all grow at the same rate, but
differ in a small constant factor, see Fig.~4 in~\cite{janus:09b}).
As expected, there is an enveloping curve for the data at
different $T$. In particular, the curves for $T=0.55, 0.625, 0.7$ appear free
from the influence of the critical point.

Similarly,$\theta(T,\xi_{12})=3 -
\dd \log I_2 / \dd \log \xi_{12}$, which we can compute numerically (see Appendix~\ref{sec:josephson}),
turns out to be a function of $x$, see the collapsing curve
Fig.~\ref{fig:replicon}. We are interested in estimating $\theta(x)$
at the experimentally relevant scale of $\xi_\text{films}=38$ for thin
films, recall our discussion of~\eqref{eq:H}.  As discussed, the RSB
and droplet pictures have diverging expectations for $\theta(0)$, that
is, for the $\xi_{12}\to\infty$ limit, so we can use them as upper and
lower bounds.  In the RSB theory, see Appendix~\ref{sec:josephson} and Fig.~\ref{fig:replicon},
we can compute an extrapolation to $\theta(0)\approx 0.30$, although
we take $\theta^\text{upper}=0.35$ as our upper bound for $\theta$.
In the droplet description, we expect $\theta(x) = C x^{\zeta}$, where
$\zeta$ is in principle the stiffness exponent $\zeta\approx
0.24$~\cite{boettcher:04}. As in~\cite{janus:10}, we find that the
droplet behavior can be reached in the infinite-$\xi_{12}$ limit, but
only with a smaller exponent $\zeta\approx0.15$, which, furthermore,
is highly sensitive to the fitting range.  Using the droplet
extrapolation for $\xi_\text{films}=38$ we obtain
$\theta(\xi_\text{films})\approx0.28$. Since our microscopically
determined $\xi_{12}$ may differ by a small constant factor from a
macroscopic measurement of $\xi$~\cite{janus:17b} we have also
considered $\xi_\text{films}=76$, which brings the exponent down to
$\theta(\xi_\text{films})=0.25$ (see Fig.~\ref{fig:replicon}).  In
short, as observed in previous work~\cite{janus:10,janus:10b}, for the
experimentally relevant scale the physics is well described by a
non-coarsening picture, with $0.25<\theta(\xi_\text{films})<0.35$
depending on the theory we use to extrapolate the data and the exact
value chosen for the experimental scale.
\begin{figure}[t]
\includegraphics[width=\linewidth]{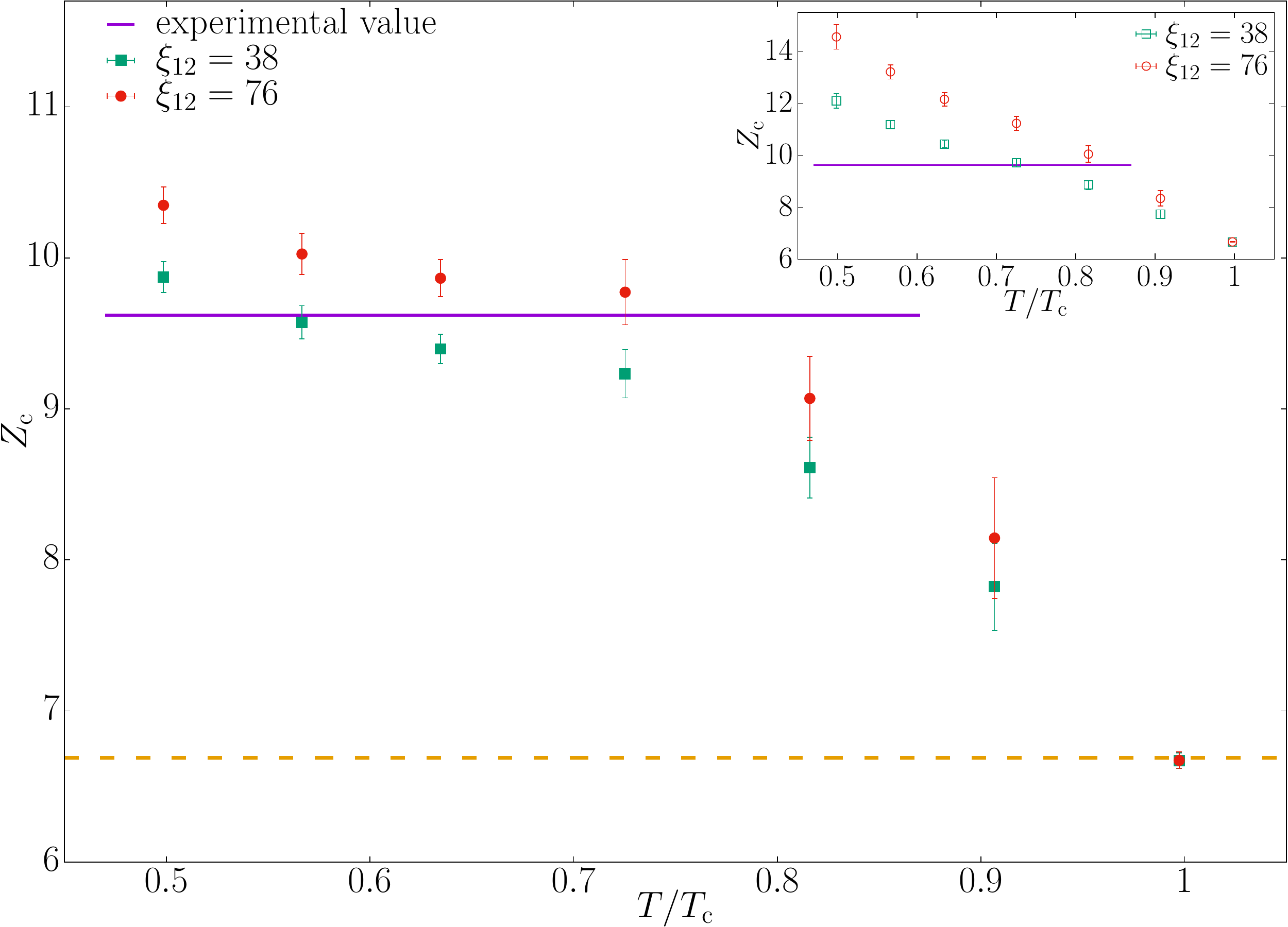}
\caption{Value of the experimental aging rate
for SGs $Z_\text{c}(T) = z(T,\xi_\text{films}) T/\Tc$,
extrapolated from our data for values of the coherence length
corresponding to thin  CuMn films.
The main plot considers an ansatz~\eqref{eq:convergente} with 
a finite $z(T,\xi_{12}\to\infty)$, which agrees very well 
with the experimental value of $Z_\text{c}(T)\approx9.62$~\cite{zhai:17}, 
indicated by the straight line, whose width represents the experimental
temperature range.
Notice that critical effects are only visible for $T>0.7$.
\emph{Inset:} Same plot but now considering a crossover 
to activated dynamics~\eqref{eq:divergente}, as in~\cite{bouchaud:01}.
This is less successful at reproducing the roughly constant $Z_\text{c}(T)$
observed in experiments.
\label{fig:zc}}
\end{figure}

As discussed
in the introduction, experiments observe a constant
$z(T) T/\Tc\approx 9.62$~\cite{zhai:17}. In the
previous discussion, on the other hand, we have found $z_\text{c}=6.69(6)$
and a growing $z(T,\xi_{12})$. Therefore, in order
to compare our results with experiments, the first step is finding
some way to extrapolate for $\xi_{12} = \xi_\text{films}$. The most
natural possibility, given the smoothness of the data in Fig.~\ref{fig:xi},
is to assume that $\tw = A(T) \xi_{12}^{z(T,\xi_{12})}$, with
a $z(T,\xi_{12})$ that tends to a finite $z_\infty(T)$ when
$\xi_{12}\to\infty$, $z(T,\xi_{12})-z_\infty(T)\propto \xi_{12}^{-\omega}$,
thus
\begin{equation}\label{eq:convergente}
\log \tw = D(T)  + z_\infty(T)\, \log\xi_{12} + E(T)\, \xi_{12}^{-\omega}\,,
\end{equation}
where $\omega$ is the exponent that controls finite-$\xi_{12}(\tw)$
corrections. At \Tc, we expect $\omega=1.12(10)$~\cite{janus:13,lulli:15,fernandez:15}.  For
$T<\Tc$, the leading behavior is given by $\omega=\theta$
(\cite{janus:10b} and section~4.2 in~\cite{janus:10}). When
fitting to~\eqref{eq:convergente}, in principle one must 
consider possible systematic effects from the fitting range
$\xi_{12}\geq\xi_{12}^\text{min}$ and the increased statistical
error due to our uncertainty in the value of $\theta$.
However, see Appendix~\ref{sec:parameters}, these effects have little impact on our 
final estimates.

An alternative interpretation is to consider a crossover to activated dynamics,
as proposed by the Saclay group~\cite{bouchaud:01,berthier:02}. Free-energy
barriers are considered from a dynamical point of view, with a growth
exponent $\varPsi$,
\begin{equation}\label{eq:divergente}
\log\tw = F(T) + z_\text{c} \log\xi_{12} + G(T) \xi_{12}^\varPsi,
\end{equation}
hence
$z(T,\xi_{12})=\mathrm{d}\log\tw/\mathrm{d}\log\xi_{12}=z_\text{c}+G(T)\varPsi\xi_{12}^\varPsi$.
Eq.~\eqref{eq:divergente}
is a refinement of droplet theory~\cite{fisher:86} and has been used
before in experiments~\cite{schins:93} and
simulations~\cite{rieger:93} with values of $\varPsi\approx
1$~\footnote{$G(T)$ in~\eqref{eq:divergente} goes to zero at \Tc as
$(\Tc-T)^{\Psi\nu}$, which is another form of the Josephson
scaling.}.  RSB theory is neutral with respect to choosing
ansatze~\eqref{eq:convergente} or~\eqref{eq:divergente}. We recall the
numerical result in infinite dimensions~\cite{billoire:01} of
$\tau\sim \exp(-N^b)$ for the time scales associated with the largest
energy barriers, with $b\approx1/3$ (see
also~\cite{rodgers:89,colborne:90}). This result can be connected with
finite $D$ at the upper critical dimensions $D_\text{u}=6$, which
yields $\varPsi(D_\text{u})\!=\!6b$.  We note, in particular,
that~\eqref{eq:convergente} can be regarded as a $\varPsi\to0$ limit
of~\eqref{eq:divergente}. With previous data it was not possible to
distinguish the behavior of~\eqref{eq:divergente} and that of a simple
power law~\cite{janus:09b}. With the present simulations, we find
that~\eqref{eq:divergente} also yields good fits for $\tw(\xi_{12})$,
with $\varPsi\approx0.4$ (again, the dependence on the fitting range
is minimal, see Appendix~\ref{sec:parameters}).

Therefore, both~\eqref{eq:convergente} and~\eqref{eq:divergente}
can explain the behavior of the data for the \emph{simulated}
scales. In order to see whether they are useful to explain 
the experiments we consider the quantity $Z_\text{c}(T) = z(T,\xi_\text{films})T/\Tc$,
where $z(T,\xi_\text{films})$ is the derivative of either~\eqref{eq:convergente}
or~\eqref{eq:divergente} at $\xi_\text{films}$. The result is plotted
in Fig.~\ref{fig:zc} (see Appendix~\ref{sec:parameters} for  the
full fit parameters). Remarkably, the convergent ansatz of~\eqref{eq:convergente}
produces an almost constant $Z_\text{c}$ in a wide $T$ range, which additionally
fits well the experimental value of $Z_\text{c} \approx9.62$. The activated
dynamics of~\eqref{eq:divergente}, on the other hand, is not a good fit for the 
experimental behavior (inset to Fig.~\ref{fig:zc}).

Using simulations for very large systems with many replicas on Janus~II we have
found that the growth of the SG coherence length is controlled by a
time-dependent $z\bigl(T,\xi(\tw)\bigr)$ exponent.  After describing the
dynamics as governed by a crossover between a critical and a low-temperature
fixed point, we have been able to model this growth quantitatively and to
extrapolate to experimental length scales. The resulting exponent is consistent
with the most recent experimental measurements for power-law dynamics. In
addition, we find clear evidence of non-coarsening dynamics at the experimental
scale and find that temperatures $T\lesssim0.7$ are free of critical effects
and therefore safe for numerical studies of the SG phase.

An open question concerns the generality of these results. Indeed,
CuMn is a Heisenberg, rather than Ising, SG. However, even the purest
Heisenberg system has unavoidable anisotropies, such as
Dzyaloshinsky-Moriya interactions~\cite{bray:82}. These interactions,
though tiny, extend over dozens of lattice spacings, which magnifies
their effect. In fact, we know that Ising is the ruling universality
class in the presence of coupling anisotropies~\cite{baityjesi:14}. We also
remark that high-quality measurements on GeMn are excellently fit with
Ising scaling laws~\cite{guchhait:17}. Our results also match the
most recent and accurate measurements on CuMn~\cite{zhai:17}.

More generally, this study is a clear demonstration of the importance 
of high-precision results for the investigation of glassiness.
Indeed, reducing the errors has shown that the aging rate slows down
during the dynamics, contrary to previous findings. A similar
change of paradigm might happen for structural glasses.

\begin{acknowledgments}
We thank R. Orbach for encouraging discussions.
 This work was partially supported by Ministerio de  Econom\'ia, Industria  y
Competitividad  (MINECO) (Spain)  through Grants No.
FIS2013-42840-P, No. FIS2015-65078-C2, No. FIS2016-76359-P, and No.
TEC2016-78358-R, by the Junta de Extremadura (Spain) through Grant No.
GRU10158 (partially funded by FEDER) and by  the DGA-FSE (Diputaci\'on General
de Arag\'on -- Fondo Social  Europeo).  This project  has received funding
from the European  Research Council (ERC)  under the European Union's Horizon
2020 research and  innovation program (Grant Nos. 694925 and 723955 -
GlassUniversality).  D. Y.  acknowledges  support by the Soft and Living
Matter Program at Syracuse University. 
\end{acknowledgments}

\appendix

\begin{figure}[tb]
\includegraphics[width=\linewidth]{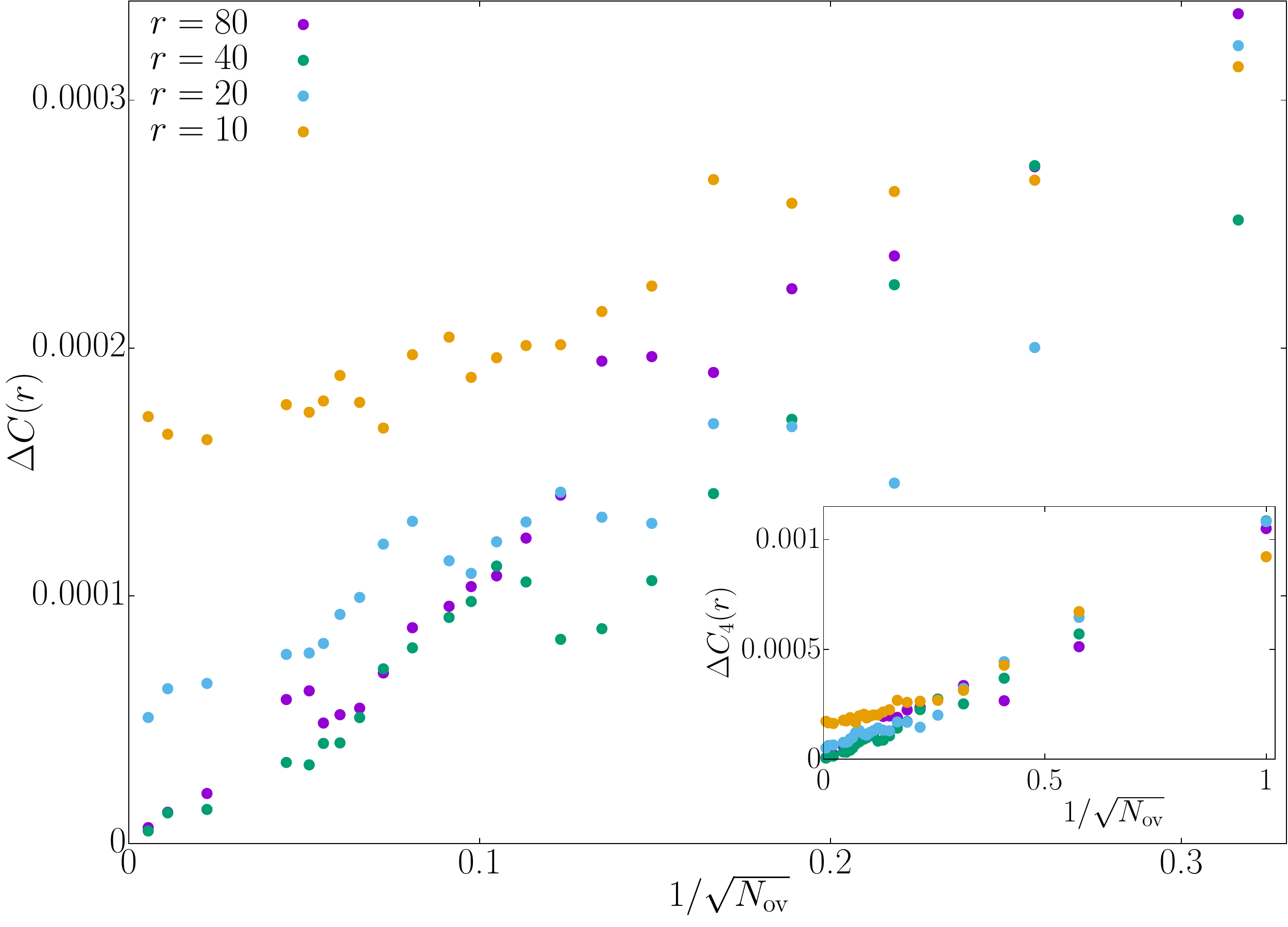}
\caption{Rapid reduction in the statistical error for 
the overlap autocorrelation function $C_4(T,r,\tw)$
with increasing number of replicas $N_\text{R}$.
We plot $\Delta C(r)$ for $T=0.7$ and $\tw=2^{32}$ and several values of
$r$ as a function of the number of possible overlaps
[$N_\text{ov} = N_\text{R}(N_\text{R}-1)/2$]. For large
values of $r$ the error is essentially linear in $1/N_\text{R}$.
The inset shows the whole range from $N_\text{R}=2$ (the minimum
to define $C_4$), while the main plot is a close up
of the large-$N_\text{R}$ sector. The simulations
reported in this paper have $N_\text{R}=256$.
\label{fig:error-NR}}
\end{figure}
\begin{figure}[hbt]
\includegraphics[width=\linewidth]{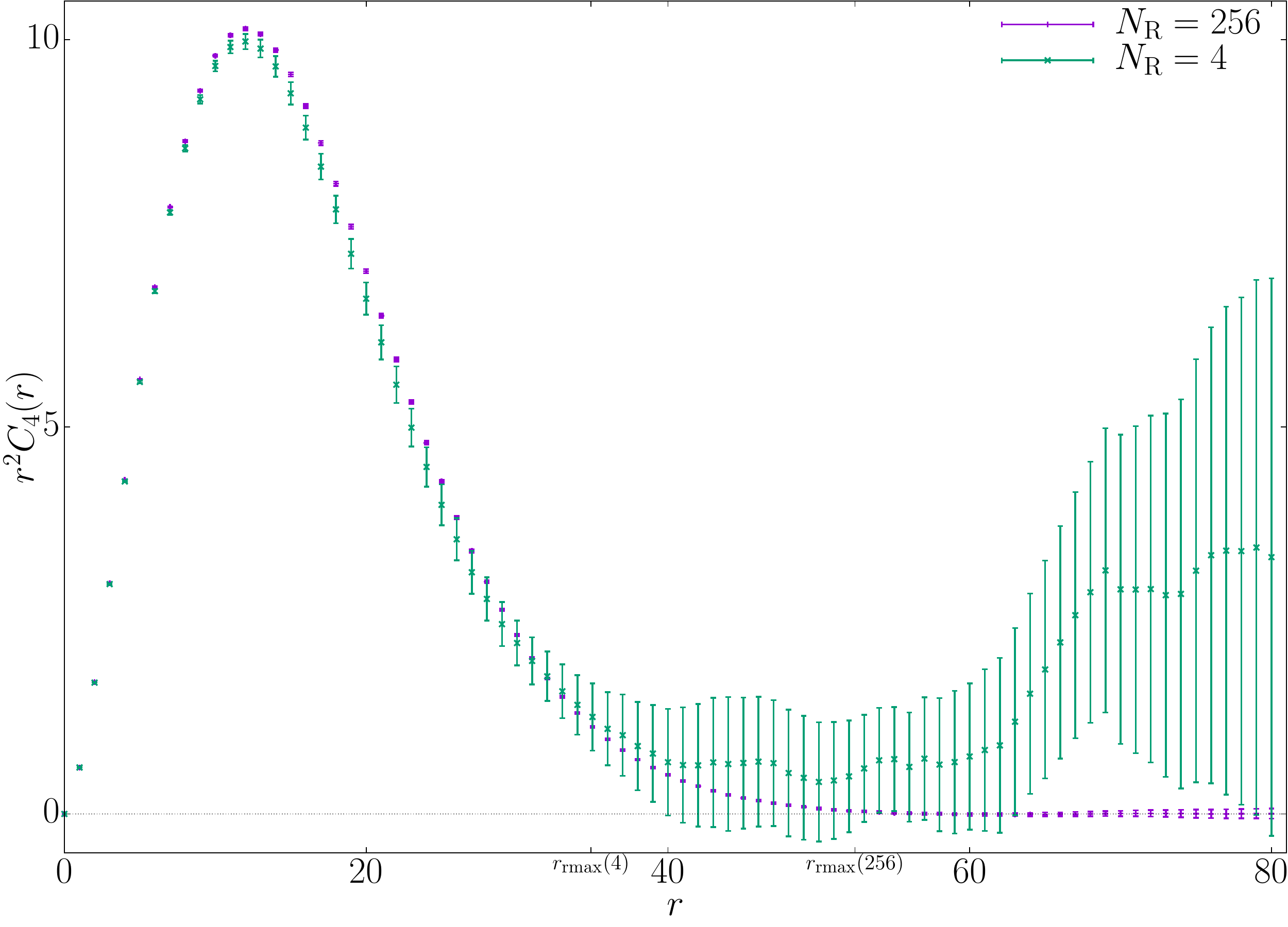}
\caption{Comparison of $r^2 C_4(r)$, function whose integral
is used to estimate the coherence length, as computed
with $4$ and $256$ replicas and the same number $N_\text{S}=16$
of samples ($T=0.7, \tw=2^{34}$). We have marked on the $x$ axis the resulting 
self-consistent cutoffs (see Appendix~\ref{sec:finite_size_effects}) used to estimate the $I_k$ integrals in
both cases.
\label{fig:error-r2Cr}}
\end{figure}

\section{Error reduction for high number of replicas}\label{sec:Nr}
As mentioned in the main paper, the choice of the number of replicas
($N_\text{R}$) and samples ($N_\text{S}$)  was taken with the aim of improving
the estimation of observables related to temperature chaos in future work,
where it is important to maximize the number of possible overlaps (pairs of replicas) 
$N_\text{ov}=N_{\mathrm{R}}(N_{\mathrm{R}} - 1)/2$.

Unexpectedly, this has led to a dramatic increase in precision.
Fig.~\ref{fig:error-NR} shows the reduction of the statistical error
in the correlation function $C_4$ as a function of $1/\sqrt{N_\text{ov}}$.
Moreover, this effect is enhanced as $r$ increases,
which leads to a qualitative improvement
in the computation of the $I_k(T,r,\tw)$ integrals (Fig.~\ref{fig:error-r2Cr}).

\begin{figure}[t]
\includegraphics[width=\linewidth]{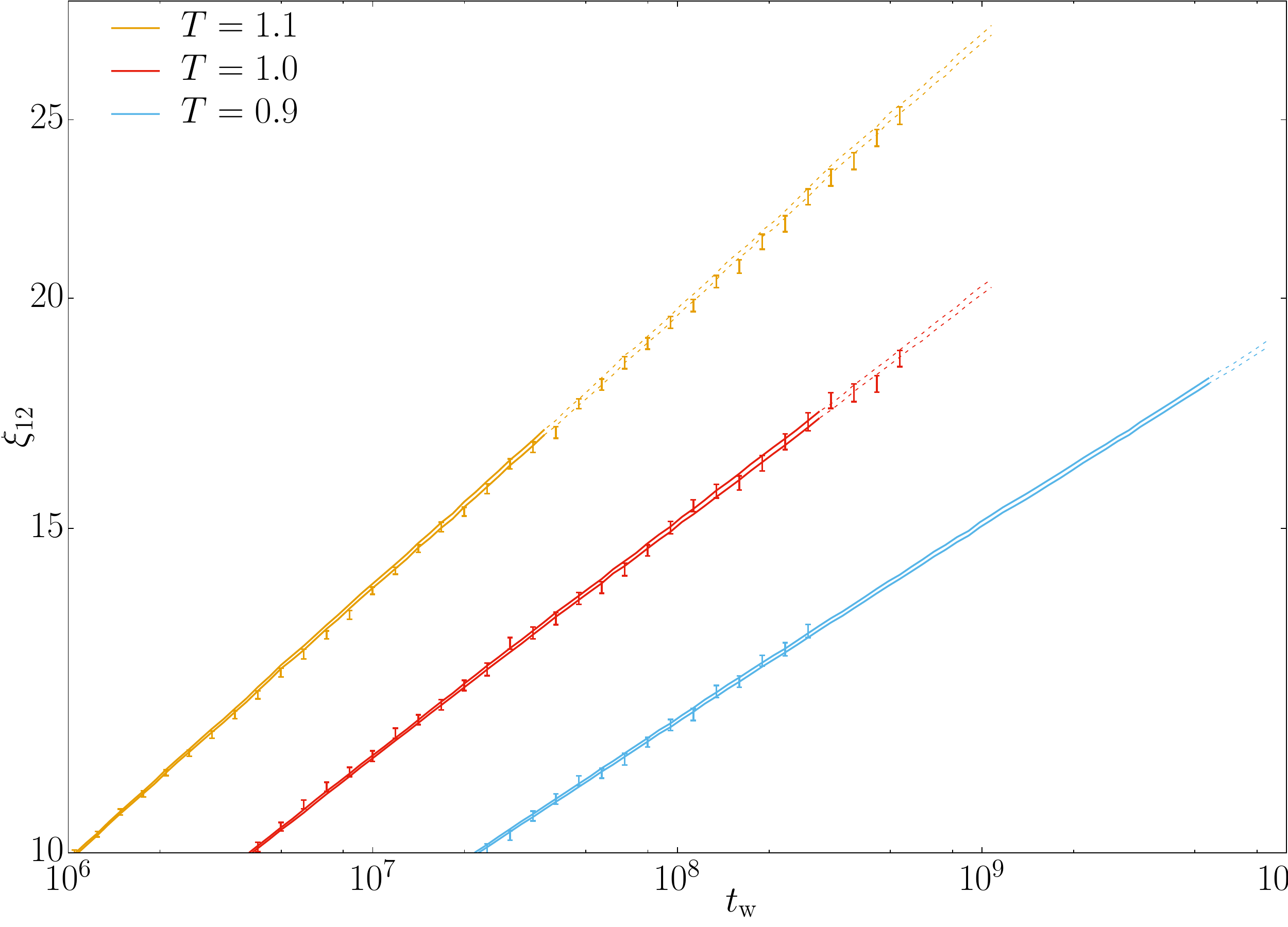}
\caption{Comparison between the $\xi_{12}(T,\tw)$ reported in this paper, 
computed in $L=160$ lattices with $N_\text{S}=16$ samples and $N_\text{R}=256$
replicas, and that of~\cite{fernandez:15} ($L=256$; $N_\text{S}=50$; $N_\text{R}=4$ 
for $T=0.9,1.0$ and $N_\text{R}=8$ for $T=1.1$). For the $L=160$ simulations, 
we plot with two parallel lines the error interval for $\xi_{12}(T,\tw)$. 
Only the \tw range depicted with  continuous lines is used in the paper, the 
extension with dashed lines represents the discarded times with
$\xi_{12}>\xi_{12}^\text{max}$ (see Table~\ref{tab:selected_ximax}).
These curves were generated with the $I^3_k$ estimator for the 
integrals~\eqref{eq:I3}.
The values from the $L=256$ simulations are plotted with conventional error bars.
Notice that both curves are compatible even beyond this cutoff.
\label{fig:xi256}}
\end{figure}

\section{Controlling finite-size effects} \label{sec:finite_size_effects}
In order to obtain an  estimate for $\xi_{k,k+1}(T,\tw) =
I_{k+1}(T,\tw)/I_k(T,\tw)$ we
need to compute the integrals
\begin{equation}\label{eq:I_SI}
I_k(T,\tw) = \int_0^\infty \mathrm{d}r\ r^k C_4(T,r,\tw) \, .
\end{equation}
As discussed in~\cite{janus:09b}, the main difficulty 
in this computation is handling the large-$r$ tail
where relative errors  [$\Delta C_4(T,r,\tw) / C_4(T,r,\tw)$]
are big. We have to consider two issues: a) how to minimize 
the statistical errors and b) how to check for finite-size effects 
(which will appear when $\xi/L$ becomes relatively large).

As explained in more detail in~\cite{janus:09b,fernandez:18a},
our estimate of $I_k(T,\tw)$ is the sum of the numerical integral
of our measured $C_4(T,r,\tw)$ up to a self-consistent
cutoff and a tail contribution estimated with a smooth extrapolating
function $F(r) \sim r^{-\theta} f\bigl(r/\xi)$.

In short, the procedure is
\begin{enumerate}
\item Obtain $F(r)$ with fits of $C_4$ in a
self-consistent region $[r_\text{min},r_\text{max}]$ 
where the signal-to-noise ratio is still good.
\item Integrate $C_4$ numerically up to some cutoff and  add
the analytical integral of $F(r)$  beyond the
cutoff to estimate the tail contribution.
\end{enumerate}
There are several choices as to how to implement these
steps, which we have used to control for systematic
effects. For the extrapolating function $F(r)$, we consider first
\begin{equation}\label{eq:F1}
F_1(r) = A_1 r^{-\theta} \ee^{-(r/\xi)^{\beta_1}}.
\end{equation}
Where $\theta$ is the replicon exponent discussed in 
the text and we fit for $A_1,\beta_1$ and $\xi$.
This analytical
form is motivated by the fact that this is the simplest choice
that avoids a pole singularity in the Fourier transform of
$C_4(T,r,\tw)$ at finite $\tw$. 
In order to check for finite-size effects we also consider
a second function $F_2(r)$ resulting from fits that
include the first-image term:
\begin{equation}
F_2^*(r) = A_2 \left[ \frac{ \ee^{-(r/\xi)^{\beta_2}}}{r^\theta} + \frac{ \ee^{-((L-r)/\xi)^{\beta_2}}}{(L-r)^\theta}\right],
\end{equation}
so we have a second extrapolating function
\begin{equation}\label{eq:F2}
F_2(r) = A_2 r^{-\theta} \ee^{-(r/\xi)^{\beta_2}}.
\end{equation}
For these fits we used $\theta=0.35$.
 However, this value has very little effect on the final computation
of $\xi_{k,k+1}$. We have checked this by recomputing the integrals with 
$\theta = 0 $ and $\theta = 1+\eta\approx 0.61$ 
(prediction of the droplet theory and influence of the $T=T_\mathrm{c}$
fixed point respectively). The different choices of $\theta$ 
led to a systematic effect smaller than $20 \%$ of the error bars 
in the worst case. 

Once we have our two extrapolating functions $F_1$ and $F_2$ we 
can combine them with the $C_4$ data in several ways:
\begin{eqnarray}
I^1_k &=& \int_0^{r_{\mathrm{max}}} dr~r^k C_4(T,r,\tw) + \int_{r_{\mathrm{max}}}^\infty dr~r^k F_1(r) \, , \\
I^2_k &=& \int_0^{r_{\mathrm{max}}} dr~r^k C_4(T,r,\tw) + \int_{r_{\mathrm{max}}}^\infty dr~r^k F_2(r) \, , \\
I^3_k &=& \int_0^{r_{\mathrm{min}}} dr~r^k C_4(T,r,\tw) + \int_{r_{\mathrm{min}}}^\infty dr~r^k F_2(r) \, . \label{eq:I3}
\end{eqnarray}

The difference between $I^2_k$ and $I^1_k$ is always under $1~\%$ in the error,
so choosing between them has no effect in any computation.
In contrast, $I^1_k$ and $I^3_k$ present measurable differences for
long $\tw$, at least for our highest temperatures, where the faster 
dynamics allows us to reach higher values of $\xi_{12}/L$.
As a (very conservative) cutoff we have discarded all the $\tw$
where $|I^1-I^3|$ is larger than $20~\%$ of the error bar,
thus obtaining a $\xi_{12}^\text{max}(T)$ below which we are 
assured not to have any finite-size effects in our $L=160$ 
systems. The reader can find the values in table~\ref{tab:selected_ximax}.

\begin{table}[t]
\centering
\begin{ruledtabular}
\begin{tabular}{cccccccc}
$T$ & $0.55$ & $0.625$ & $0.7$ & $0.8$ & $0.9$ & $1.0$ & $1.1$  \\
$\xi_\mathrm{max}$ & - & - & - & - & $18.1$ & $17.3$ & $17$ \\
\end{tabular}
\end{ruledtabular}
\caption{Cutoff values of $\xi_{12}^\mathrm{max}(T)$  below which we are 
guaranteed no finite-size effects in our $L=160$ lattices. For $T<0.9$
the growth of $\xi_{12}$ is very slow and we never reach the cutoff value.
\label{tab:selected_ximax}}
\end{table}

As a final check that our data is not affected by finite-size effects, we have
compared our $\xi_{12}(T,\tw)$ with that of~\cite{fernandez:15}. This reference
considers shorter simulations but with $L=256$ and $50$ samples. As 
shown in Fig.~\ref{fig:xi256}, the $L=160$ and $L=256$ data coincide even 
beyond our cutoff $\xi_{12}^\text{max}$.

\begin{figure}[hbt]
\includegraphics[width=\linewidth]{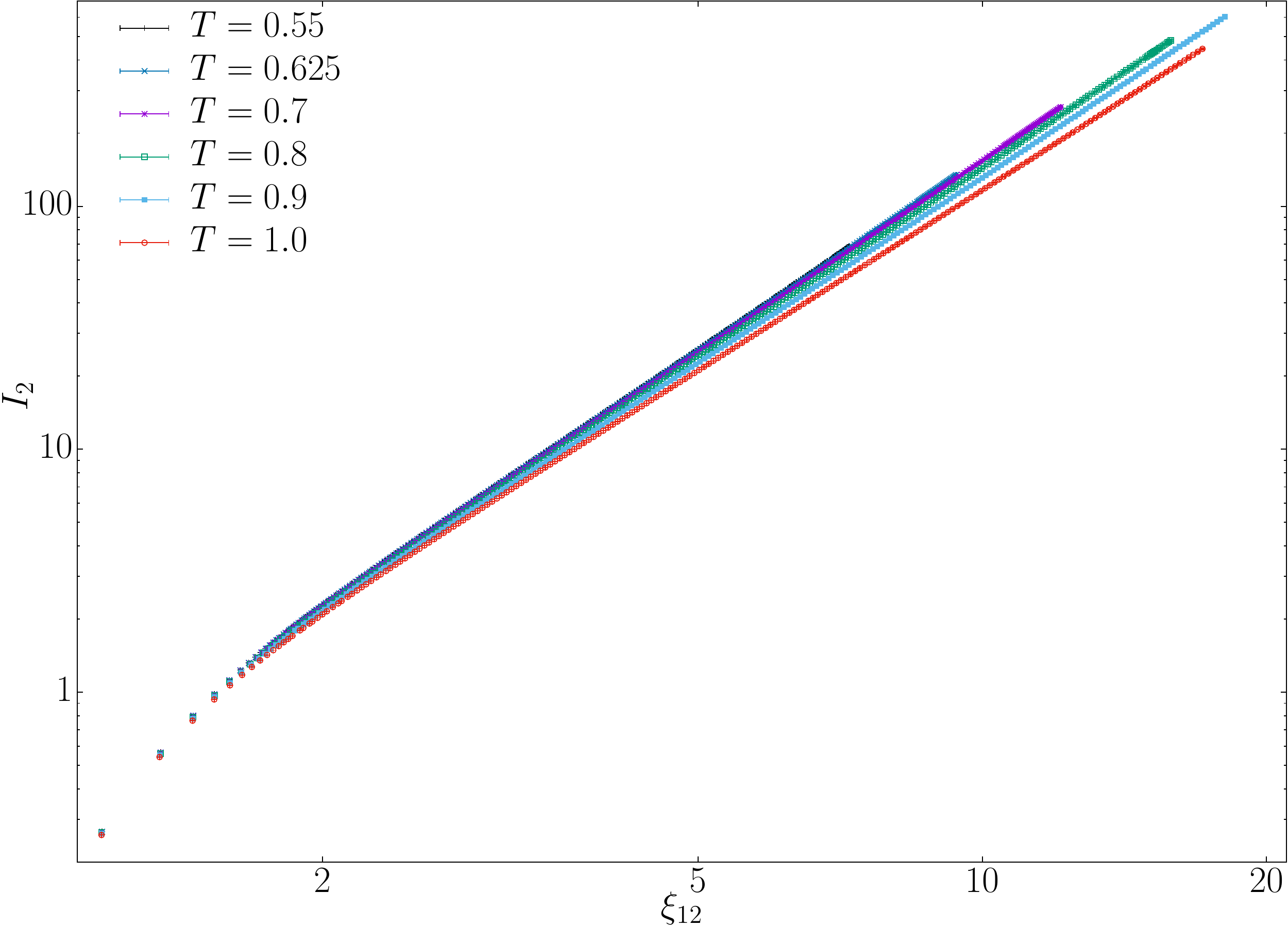}
\caption{Integral $I_2$ as a function of $\xi_{12}$ 
in a logarithmic scale, for all our $T<\Tc$ temperatures.
We use the numerical derivative of this curve
to compute the replicon exponent $\theta$.
\label{fig:I2xi}}
\end{figure}
\begin{figure}[hbt]
\includegraphics[width=\linewidth]{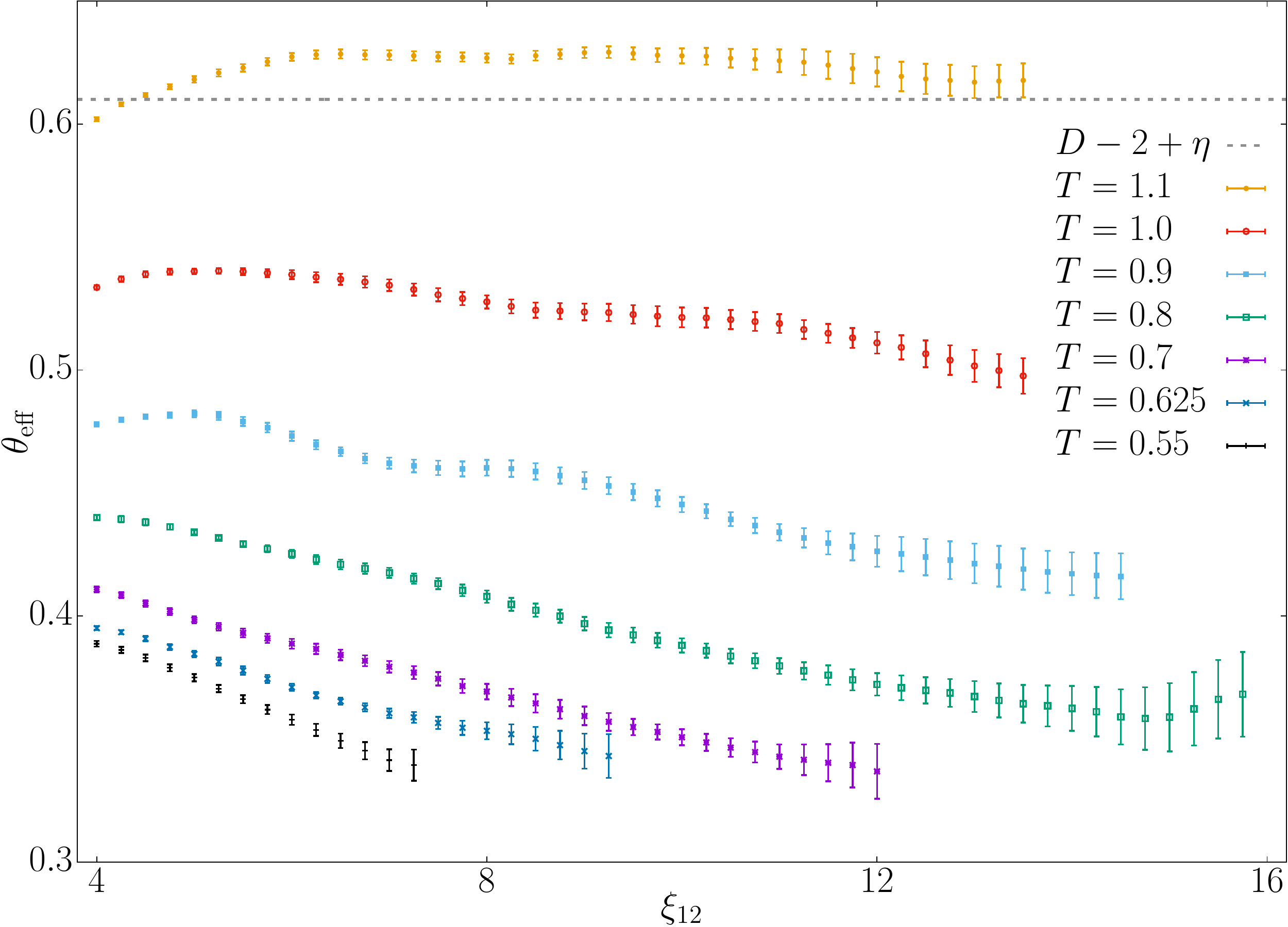}
\caption{Value of the replicon exponent $\theta(T,\xi_{12})$ computed
from a numerical derivative of $\log I_2$ as a function of $\log\xi_{12}$,
nicely illustrating the crossover between the $T=\Tc$ and 
$T=0$ fixed points. 
\label{fig:replicon_sin_reescalar}}
\end{figure}

\section{Josephson Crossover}\label{sec:josephson}
In this section we will give additional details on the Josephson
crossover which describes how $C_4(T,r,\tw)$ changes from being
dominated by the $T=\Tc$ fixed point 
to being dominated by the $T=0$ behavior as $T$ and $\tw$ vary.  
Assuming that $\xi(T,\tw) \gg  \ell_\text{J}(T) \sim (T-\Tc)^{-\nu}$, the
crossover takes
the form:
\begin{equation}
  C_4(T,r,\tw) \sim
  \begin{cases}\displaystyle
    \frac{1}{r^{D-2+\eta}}\,,  & r\ll \ell_\text{J}(T)\,, \\[3mm]
\displaystyle     \frac{\ell_\text{J}^\theta}{\ell_\text{J}^{D-2+\eta}} \frac{1}{r^\theta} f_\text{cutoff}(r/\xi)\,, & r \gg
\ell_\text{J}(T) \,.
    \end{cases}
\end{equation}
In this equation, $f_\text{cutoff}(x)$ is an analytical function decaying as $\exp(-x^\beta)$.
The prefactor for $\ell_\text{J}^\theta/\ell_\text{J}^{D-2+\eta}$ is fixed by the condition 
that the two asymptotic expansions connect smoothly at $r=\ell_\text{J}$.
We arrive at an asymptotic expansion for the $I_k$ integrals:
\begin{equation}
I_k(T,\xi) = \frac{F_k}{\ell_\text{J}^{D-2+\eta}} \left(\frac{\xi}{\ell_\text{J}}\right)^{k+1-\theta}
\left[ 1 + a_k\left(\frac{\xi}{\ell_\text{J}}\right)^{k+1-\theta}+\ldots\right]\,,
\end{equation}
where $F_k$ and $a_k$ are amplitudes.
Finally, we need to eliminate the unknown $\xi$ in favor of the computable $\xi_{12}$,
\begin{equation}
\xi_{12}(T,\xi) = \frac{F_2}{F_1} \xi \left[ 1+ a_1'\left(\frac{\xi}{\ell_\text{J}}\right)^{2-\theta}
+  a_2\left(\frac{\xi}{\ell_\text{J}}\right)^{3-\theta}+\ldots\right]\,,
\end{equation}
where $a_1'$ considers contributions both from the numerator ($-a_1$) and 
from the denominator.  The easiest way to obtain $\theta$ is to study the evolution
of  $\log I_2$ as a function of $\log \xi$. However, we have to settle
for using $\log\xi_{12}$ as independent variable (see Fig.~\ref{fig:I2xi}).

We can define an effective $\theta(T,\xi_{12})$ as
\begin{align}
\theta(T,\xi_{12}) &= 3 - \frac{\dd \log I_2(T,\xi_{12})}{\dd \log \xi_{12}}\\
		   &= \theta + b_2
\left(\frac{\xi_{12}}{\ell_\text{J}}\right)^{\theta-2}+b_3
\left(\frac{\xi_{12}}{\ell_\text{J}}\right)^{\theta-3}+\ldots\, . \label{eq:fit}
\end{align}
To estimate this derivative for a given $\xi_{12}^*$, we 
fit $\log I_2$ to a quadratic polynomial in $\log \xi_{12}$ 
in a $[0.75\xi_{12}^*,1.25\xi_{12}^*]$ window. We then 
take the derivative of this polynomial at $\xi^*$. The procedure,
as well as the wiggles in the resulting values of $\theta$ due to the 
extreme data correlation (see Fig.~\ref{fig:replicon_sin_reescalar})
may remind the reader of Fig.~1 in~\cite{janus:08b}.

We have computed a fit to the first two terms in~\eqref{eq:fit}  
in the range $0\leq \ell_\text{J}/\xi_{12}\leq 0.33$, resulting 
in the value of $\theta\approx 0.30$ reported in the main text.

The previous analysis solves the problem of the crossover between
the $T=\Tc$ and $T=0$ fixed points. However, in the framework of the
droplet picture one would also need to consider corrections to scaling
at the $T=0$ fixed point. This is precisely what the droplet fit in the 
main text to $\theta(x) \simeq Cx^\zeta$ does.

\section{Parameter choices in our fits}\label{sec:parameters}
We will discuss separately the choice of $\xi_{\mathrm{min}}$ for different
temperatures and the choice of the value of $\omega$.

\subsection{ Selection of \boldmath $\xi_{12}^{\mathrm{min}}$ for each temperature}
We have reported fits of our data to three different functional forms

\begin{align}
\log \tw&= c_0(T) + c_1(T) \log \xi_{12} + c_2(T) \log^2 \xi_{12},\\
\log t_\mathrm{w} &= D(T) + z_\infty(T) \log \xi_{12} + E(T) \xi_{12}^{-\omega} \, , \label{eq:RSB}\\
\log t_\mathrm{w} &= F(T) + z_\mathrm{c} \log \xi_{12} + G(T) \xi_{12}^{\varPsi} \, .
\label{eq:Bouchaud}
\end{align}

In these fits we have used $z_\text{c}=6.69$ and $\omega=0.35$ ($T<\Tc$),
$\omega=1.12$ ($T=\Tc$), as discussed in the main text. 
Full results for the fits to~\eqref{eq:RSB} and~\eqref{eq:Bouchaud}
can be seen in tables~\ref{tab:RSB_omega_0.35} and~\ref{tab:Saclay_6.69}, 
for different fitting ranges. We include for both cases the extrapolated
values of $z(T,\xi)$ for the experimental scale
(as explained in the main text we use both $\xi_{12}=38$ and $\xi_{12}=76$) and
for~\eqref{eq:RSB}
also the value of the $\xi\to\infty$ aging rate $z_\infty$.

In order to make the choice of fitting range for the values 
plotted in the paper we have followed two criteria.
Firstly we require the parameters
of the fit to be stable inside the error when we increase $\xi_{12}^\mathrm{min}$.
Secondly, we impose that $\xi_\mathrm{min}$ be monotonically increasing in $T$ (with the 
exception of \Tc, which has different behavior).
Table~\ref{tab:selected_ximin} shows our final choices for $\xi_{12}^\text{min}(T)$, 
which is the same for all three fits.

\begin{table}[htb]
\centering
\begin{ruledtabular}
\begin{tabular}{cccccccc}
$T$ & $0.55$ & $0.625$ & $0.7$ & $0.8$ & $0.9$ & $1.0$ & $1.1$  \\
$\xi_{12}^\mathrm{min}$ & $4$ & $5$ & $6$ & $8$ & $8$ & $9$ & $5$ \\
\end{tabular}
\end{ruledtabular}
\caption{Values of $\xi_{12}^\text{min}(T)$ determining
the common fitting range $\xi_{12}\geq\xi_{12}^\text{min}$   
for our three different fits of $\log\tw$ as a function of
$\log \xi_{12}$. 
\label{tab:selected_ximin}}
\end{table}

\subsection{Selection of \boldmath $\omega$}
For our most important result, namely the extrapolation of the aging
rate to the experimental scale of $\xi_{12}=38,76$, we have repeated
our fits with our upper and lower bounds for $\omega=\theta(\xi_\text{films})$
(RSB and droplet extrapolations, respectively).  The results are completely
compatible, as we can see in table~\ref{tab:omega_xi38}.

\begin{table*}[p]
\begin{ruledtabular}
\begin{tabular}{lcccccccc}
& & $\xi_{\mathrm{min}}$= 3.5 & $\xi_{\mathrm{min}}$= 4 & $\xi_{\mathrm{min}}$= 5 & $\xi_{\mathrm{min}}$= 6 & $\xi_{\mathrm{min}}$= 7 & $\xi_{\mathrm{min}}$= 8 & $\xi_{\mathrm{min}}$= 9 \\ \hline
\multirow{4}{*}{{$T=0.55$}}&
$z_\infty$ & 23.61(28)   & \bfseries 24.22(40)   & 25.30(86)   & 22.9(31)   & & &   \\ 
&$z (\xi\!=\!38)$ & 19.49(15)   &  \bfseries 19.80(20)   & 20.32(41)  & 19.2(14)   & & &   \\
&$z (\xi\!=\!76)$ & 20.38(18)   &  \bfseries 20.75(24)   & 21.39(51)   & 20.0(18)   & & &   \\
&$\chi^2$/dof &  $40(17)/133$ &  \bfseries 10.2(54)/111 & 3.0(12)/73 & 1.71(76)/40  &  &  & \\ 
\hline
\multirow{4}{*}{{$T=0.625$}}&
$z_\infty$ & 19.85(17)   & 20.26(23)   &  \bfseries 20.60(41)   & 20.16(84)   &  &  &  \\ 
&$z (\xi\!=\!38)$ & 16.538(91)   & 16.74(12)   &  \bfseries 16.90(19)   & 16.71(37)   &  &  &  \\
&$z (\xi\!=\!76)$ & 17.25(11)   & 17.50(14)   &  \bfseries 17.69(24)   & 17.45(47)   &  &  &  \\
&$\chi^2$/dof & 81(34)/167
 &  18(10)/147
 &  \bfseries  8.3(21)/114
&  5.1(19)/86
&  &  &  \\ \hline
\multirow{4}{*}{{$T=0.7$}}&
$z_\infty$ & 17.04(18)   & 17.23(21)   & 17.61(27)   &  \bfseries 18.23(35)   & 18.63(62)   &  &  \\ 
&$z (\xi\!=\!38)$ & 14.295(88)   & 14.38(11)   & 14.55(13)   &  \bfseries 14.81(15)   & 14.96(25)&  &  \\
&$z (\xi\!=\!76)$ & 14.89(11)   & 15.00(13)   & 15.21(16)   &  \bfseries 15.54(19)   & 15.75(32)   &  &  \\ 
&$\chi^2$/dof & 116(40)/190 & 66(36)/173 & 33(24)/144 &  \bfseries 9.3(84)/119 & 4.9(21)/98
 &  &  \\ \hline
\multirow{4}{*}{{$T=0.8$}}&
$z_\infty$ & 13.76(15)   & 14.06(19) & 14.53(26)   & 15.19(35)   & 15.68(42)   &  \bfseries 16.18(58)   & 16.55(78)   \\ 
& $z (\xi\!=\!38)$ & 11.787(73)   & 11.921(89)   & 12.11(12)   & 12.37(15)   & 12.55(17)   & \bfseries  12.73(22)   & 12.85(28)   \\
& $z (\xi\!=\!76)$ & 12.211(93)   & 12.38(11)   & 12.63(15)   & 12.98(19)   & 13.23(23)   & \bfseries  13.47(30)   & 13.65(39)   \\ 
& $\chi^2$/dof & 351(104)/185 & 188(72)/170& 93(41)/146& 27(16)/125& 12.4(82)/107
& \bfseries  6.2(31)/91
& 4.9(21)/77
\\ \hline
\multirow{4}{*}{{$T=0.9$}}&
$z_\infty$ & 11.00(13)  & 11.29(18)   & 11.54(24)   & 11.80(31)   & 12.55(41)   & \bfseries  13.16(68)   & 12.3(13)   \\ 
& $z (\xi\!=\!38)$ & 9.748(65)   & 9.883(93)   & 9.98(11)   & 10.08(13)   & 10.34(16)   & \bfseries  10.55(25)   & 10.33(41)   \\
& $z (\xi\!=\!76)$ & 10.017(82)   & 10.18(11)   & 10.32(14)   & 10.45(17)   & 10.82(21)   &  \bfseries 11.11(34)   & 10.80(60)   \\ 
& $\chi^2$/dof & 310(150)/165 & 129(64)/152 & 79(44)/131 & 63(35)/113 & 22(13)/98
 & \bfseries  5.9(21)/84
 & 6.4(79)/72
 \\ \hline
\multirow{4}{*}{{$T=1.0$}}&
$z_\infty$ & 8.60(11)   & 8.69(15)   & 8.83(20)   & 8.97(53)   & 9.29(45)   & 9.36(46)   &  \bfseries 10.28(89)   \\
& $z (\xi\!=\!38)$ & 8.041(59)   & 8.080(73)   & 8.132(93)   & 8.21(22)   & 8.27(18)   & 8.34(17)   &  \bfseries 8.63(32)   \\ 
& $z (\xi\!=\!76)$ & 8.162(74) & 8.210(86) & 8.28(11) & 8.38(29) & 8.46(24) & 8.57(24) & \bfseries  8.98(44)   \\ 
& $\chi^2$/dof & 43(30)/137
 & 27(18)/126
 & 16(13)/107
 & 12(25)/91
 & 10.4(91)/78
 & 8.4(60)/66
 &  \bfseries 2.9(21)/55
\\ \hline
\multirow{4}{*}{{$T=1.1$}}&
$z_\infty$ & 6.672(44) & 6.671(41) & \bfseries  6.689(63) & 6.751(84) & 6.80(12) & 7.00(18) & 7.02(21)  \\
& $z (\xi\!=\!38)$ & 6.682(32) & 6.673(41) & \bfseries  6.694(50) & 6.732(68) & 6.77(10) & 6.92(14) & 6.94(16)   \\
& $z (\xi\!=\!76)$ & 6.677(33) & 6.671(41) & \bfseries  6.691(54) & 6.742(72) & 6.79(11) & 6.96(16) & 6.99(16)   \\
& $\chi^2$/dof & 32(19)/119
 & 31(20)/109
 &  \bfseries 26(16)/92
 & 19(10)/78
 & 16.8(74)/66
 & 5.9(20)/55
 & 6.3(27)/46
\end{tabular}
\end{ruledtabular}
\caption{Parameters of the fits to~\eqref{eq:RSB} for different
fitting ranges $\xi_{12}\geq\xi_{12}^\text{min}$. 
We use $\omega = 0.35$ ($\omega = 1.12$ for $T = T_\mathrm{c}$).
The fitting range that we choose for our final values is highlighted in boldface.
\label{tab:RSB_omega_0.35}}
\end{table*}

\begin{table*}[p]
\begin{ruledtabular}
\begin{tabular}{lcccccccc}
& & $\xi_{\mathrm{min}}$= 3.5 & $\xi_{\mathrm{min}}$= 4 & $\xi_{\mathrm{min}}$= 5 & $\xi_{\mathrm{min}}$= 6 & $\xi_{\mathrm{min}}$= 7 & $\xi_{\mathrm{min}}$= 8 & $\xi_{\mathrm{min}}$= 9 \\ \hline

\multirow{4}{*}{{$T=0.55$}}& 
$z(\xi\!=\!38)$ & 24.07(41)   & \bfseries24.25(55)  & 24.6(11)& 24.7(81)  &  &  & \\
&$z(\xi\!=\!76)$ & 28.86(69)   & 2\bfseries9.18(95)  & 29.9(19)   & 30(15)  &  &  & \\
& $B(T)$ & 13.78(65)   & \bfseries13.45(92) & 12.8(18)  & 18(13) &  &  &  \\ 
& $\varPsi$ & 0.3512(92)   & \bfseries0.355(21)   & 0.372(33)   & 0.29(24)  &  &  & \\
& $\chi^2$/dof & 13.3(47)/133
 & \bfseries6.8(20)/111
 & 3.2(15)/73
 & 1.7(27)/40
 &  &  &  \\ \hline
\multirow{4}{*}{{$T=0.625$}}& 
 $z(\xi\!=\!38)$ & 19.73(22)   & 19.72(28)   & \bfseries19.36(45)   & 18.53(77)  &  &  & \\
& $z(\xi\!=\!76)$ & 23.33(38)   & 23.31(49)   & \bfseries22.66(79)   & 21.1(13)  &  &  & \\
& $B(T)$ & 10.36(37)   & 10.39(52)   & \bfseries11.3(11)   & 14.0(30)  &  &  & \\
& $\varPsi$ & 0.354(14)   & 0.352(12)   & \bfseries0.334(21)   & 0.290(39)  &  &  &  \\
& $\chi^2$/dof & 19(10)/167
 & 15(10)/147
 & \bfseries8.5(33)/114
 & 4.5(14)/86
 &  &  &  \\ \hline
\multirow{4}{*}{{$T=0.7$}}& 
$z(\xi\!=\!38)$ & 16.58(22)   & 16.44(23)   & 16.35(27)   &\bfseries 16.51(32)   & 16.55(52) &  &   \\ 
& $z(\xi\!=\!76)$ & 19.40(37)   & 19.14(40)   & 18.98(47)   &\bfseries 19.29(58)   & 19.4(10)  &  &  \\
& $B(T)$ & 7.32(34)   & 7.63(43)   & 7.84(59)   &\bfseries 7.41(75)   & 7.3(14)   &  & \\ 
& $\varPsi$ & 0.364(13)   & 0.354(12)   & 0.354(24)   &\bfseries 0.358(23)   & 0.360(39)   &  & \\
& $\chi^2$/dof & 49(38)/190
 & 28(20)/173
 & 10.5(83)/144
 & \bfseries6.3(31)/119
 & 5.7(29)/98
 &  &  \\ \hline
\multirow{4}{*}{{$T=0.8$}}& 
$z(\xi\!=\!38)$ & 13.37(18)   & 13.39(21) & 13.45(25) & 13.68(31) & 13.80(35) & \bfseries13.94(45) & 14.1(17)   \\
& $z(\xi\!=\!76)$ & 15.44(33) & 15.48(37) & 15.60(46) & 16.06(59)   & 16.31(68)   & \bfseries16.59(93)   & 17.1(38)   \\ 
& $B(T)$ & 4.16(25)   & 4.13(29)   & 4.01(38)   & 3.57(43)   & 3.36(48)   & \bfseries3.13(66)   & 3.0(19)   \\
& $\varPsi$ & 0.392(13) & 0.390(19) & 0.395(18)   & 0.421(27) & 0.443(31) & \bfseries0.447(50) & 0.46(18) \\ 
& $\chi^2$/dof & 31(19)/185
 & 29(19)/170
 & 22(17)/146
 & 10.0(60)/125
 & 7.5(34)/107
 & \bfseries5.5(21)/91
 & 5(11)/77
 \\ \hline
\multirow{4}{*}{{$T=0.9$}}& 
$z(\xi\!=\!38)$ & 10.76(17) & 10.86(21) & 10.82(24) & 10.86(27) & 11.23(35) & \bfseries11.49(54) & 11.13(56) \\
& $z(\xi\!=\!76)$ & 12.12(30) & 12.31(39) & 12.24(45) & 12.31(53) & 13.12(73) & \bfseries13.7(12)   & 12.9(12)  \\
& $B(T)$ & 2.15(19) & 2.01(23) & 2.07(31) & 2.00(39) & 1.52(31)   & \bfseries1.18(45)   & 1.67(77)  \\
& $\varPsi$ & 0.417(23)  & 0.430(34) & 0.431(28) & 0.427(41) & 0.490(51) & \bfseries0.546(88) & 0.47(10) \\
& $\chi^2$/dof & 68(44)/165
 & 46(25)/152
 & 41(25)/131
 & 38(24)/113
 & 17(10)/98
 & \bfseries8.7(45)/84
 & 4.8(25)/72
 \\ \hline
\multirow{4}{*}{{$T=1.0$}}& 
$z(\xi\!=\!38)$ & 8.53(15) & 8.54(18) & 8.55(20) & 8.56(30) & 8.59(60) & 8.74(72) &\bfseries 9.22(18)  \\
& $z(\xi\!=\!76)$ & 9.19(27) & 9.20(33) & 9.22(39) & 9.25(58) & 9.3(12) & 9.7(16) &\bfseries 10.9(45)  \\
& $B(T)$ & 0.85(14) & 0.84(18) & 0.83(22) & 0.79(40) & 0.7(10)   & 0.5(10) &\bfseries 1.4(19)  \\
& $\varPsi$ & 0.440(36)   & 0.441(51)   & 0.444(64) & 0.45(10) & 0.49(25) & 0.55(30)   & \bfseries0.34(71) \\ 
& $\chi^2$/dof & 12.6(95)/137
 & 12.1(90)/126
 & 10.0(87)/107
 & 9.3(83)/91
 & 8.2(97)/78
 & 07(11)/66
 & \bfseries11(11)/55
 \\ \hline
\multirow{4}{*}{{$T=1.1$}}& 
$z(\xi\!=\!38)$ & 6.684(12)   & 6.682(14)   &\bfseries 6.684(11)   & 6.672(31)   & 6.683(32)& 6.694(31)& 6.712(41)   \\ 
& $z(\xi\!=\!76)$ & 6.684(13)   & 6.681(11)   & \bfseries 6.682(14)   & 6.674(41)   & 6.684(41)& 6.692(31)& 6.721(42)  \\ 
& $B(T)$ & 1.9(10)   & 1.71(91) & \bfseries0.4(27) & 0.02(64)  & 0.0(26)   & 1.5(26)   & 1.2(10)   \\ 
& $\varPsi$ & -0.0030(49)   & 0.0037(68)   &\bfseries 0.03(15)   & 0.29(42)   & 0.37(55)   & 0.002(16)   & 0.023(51)   \\
& $\chi^2$/dof & 34(20)/119
 & 33(19)/109
 &\bfseries 27(18)/92
 & 25(18)/78
 & 23(15)/66
 & 21(14)/55
 & 11.0(26)/46
 \\ \hline
\end{tabular}
\end{ruledtabular}
\caption{Parameters of the fits to~\eqref{eq:Bouchaud} for different
fitting ranges $\xi_{12}\geq\xi_{12}^\text{min}$. 
We use $z_\text{c} = 6.69$.
The fitting range that we choose for our final values is highlighted in boldface.
\label{tab:Saclay_6.69}}
\end{table*}

\begin{table}[htb]
\begin{ruledtabular}
\begin{tabular}{lccccc}
& \multicolumn{2}{c}{$z(T,\xi_{12}=38)$} & & \multicolumn{2}{c}{$z(T,\xi_{12}=76)$} \\ 
       & $\omega = 0.35$    & \multicolumn{1}{c}{$\omega = 0.28$} & 
       & $\omega = 0.35$    & \multicolumn{1}{c}{$\omega = 0.25$} \\ 
\hline
$T=0.55$  & 19.80(20) & 20.08(22) &  & 20.75(24) & 21.41(27)                    \\
$T=0.625$ & 16.90(19)  & 17.07(20)& & 17.69(24)  & 18.13(27)                   \\
$T=0.7$   & 14.81(15) & 14.93(16)&  & 15.54(19) & 15.87(21)                   \\
$T=0.8$   & 12.73(22) & 12.81(23)&  & 13.47(30) & 13.71(32)                    \\
$T=0.9$   & 10.55(25) & 10.61(26)&  & 11.11(34) & 11.28(37)                   \\
$T=1.0$   & 8.63(32) & 8.68(33)  &  & 8.98(44) & 9.02(42)                  \\ 
\end{tabular}
\end{ruledtabular}
\caption{Comparison of our estimates of the experimental aging
rate $z(T,\xi_{12}=\xi_\text{films})$ for $\xi_\text{films}=38$
and $\xi_\text{films}=76$
using our lower and upper bounds for $\omega=\theta(\xi_\text{films})$.
The choice of $\omega$ is immaterial, since even in the worst case
(lowest temperatures for $\xi_\text{films}=76$) there is 
only a two-sigma difference.
\label{tab:omega_xi38}}
\end{table}


\begin{thebibliography}{45}%
\makeatletter
\providecommand \@ifxundefined [1]{%
 \@ifx{#1\undefined}
}%
\providecommand \@ifnum [1]{%
 \ifnum #1\expandafter \@firstoftwo
 \else \expandafter \@secondoftwo
 \fi
}%
\providecommand \@ifx [1]{%
 \ifx #1\expandafter \@firstoftwo
 \else \expandafter \@secondoftwo
 \fi
}%
\providecommand \natexlab [1]{#1}%
\providecommand \enquote  [1]{``#1''}%
\providecommand \bibnamefont  [1]{#1}%
\providecommand \bibfnamefont [1]{#1}%
\providecommand \citenamefont [1]{#1}%
\providecommand \href@noop [0]{\@secondoftwo}%
\providecommand \href [0]{\begingroup \@sanitize@url \@href}%
\providecommand \@href[1]{\@@startlink{#1}\@@href}%
\providecommand \@@href[1]{\endgroup#1\@@endlink}%
\providecommand \@sanitize@url [0]{\catcode `\\12\catcode `\$12\catcode
  `\&12\catcode `\#12\catcode `\^12\catcode `\_12\catcode `\%12\relax}%
\providecommand \@@startlink[1]{}%
\providecommand \@@endlink[0]{}%
\providecommand \url  [0]{\begingroup\@sanitize@url \@url }%
\providecommand \@url [1]{\endgroup\@href {#1}{\urlprefix }}%
\providecommand \urlprefix  [0]{URL }%
\providecommand \Eprint [0]{\href }%
\providecommand \doibase [0]{http://dx.doi.org/}%
\providecommand \selectlanguage [0]{\@gobble}%
\providecommand \bibinfo  [0]{\@secondoftwo}%
\providecommand \bibfield  [0]{\@secondoftwo}%
\providecommand \translation [1]{[#1]}%
\providecommand \BibitemOpen [0]{}%
\providecommand \bibitemStop [0]{}%
\providecommand \bibitemNoStop [0]{.\EOS\space}%
\providecommand \EOS [0]{\spacefactor3000\relax}%
\providecommand \BibitemShut  [1]{\csname bibitem#1\endcsname}%
\let\auto@bib@innerbib\@empty
\bibitem [{\citenamefont {Mydosh}(1993)}]{mydosh:93}%
  \BibitemOpen
  \bibfield  {author} {\bibinfo {author} {\bibfnamefont {J.~A.}\ \bibnamefont
  {Mydosh}},\ }\href@noop {} {\emph {\bibinfo {title} {Spin Glasses: an
  Experimental Introduction}}}\ (\bibinfo  {publisher} {Taylor and Francis},\
  \bibinfo {address} {London},\ \bibinfo {year} {1993})\BibitemShut {NoStop}%
\bibitem [{\citenamefont {Young}(1998)}]{young:98}%
  \BibitemOpen
  \bibfield  {author} {\bibinfo {author} {\bibfnamefont {A.~P.}\ \bibnamefont
  {Young}},\ }\href@noop {} {\emph {\bibinfo {title} {Spin Glasses and Random
  Fields}}}\ (\bibinfo  {publisher} {World Scientific},\ \bibinfo {address}
  {Singapore},\ \bibinfo {year} {1998})\BibitemShut {NoStop}%
\bibitem [{\citenamefont {Marinari}\ \emph {et~al.}(2000)\citenamefont
  {Marinari}, \citenamefont {Parisi}, \citenamefont {Ricci-Tersenghi},
  \citenamefont {Ruiz-Lorenzo},\ and\ \citenamefont {Zuliani}}]{marinari:00}%
  \BibitemOpen
  \bibfield  {author} {\bibinfo {author} {\bibfnamefont {E.}~\bibnamefont
  {Marinari}}, \bibinfo {author} {\bibfnamefont {G.}~\bibnamefont {Parisi}},
  \bibinfo {author} {\bibfnamefont {F.}~\bibnamefont {Ricci-Tersenghi}},
  \bibinfo {author} {\bibfnamefont {J.~J.}\ \bibnamefont {Ruiz-Lorenzo}}, \
  and\ \bibinfo {author} {\bibfnamefont {F.}~\bibnamefont {Zuliani}},\ }\href
  {\doibase 10.1023/A:1018607809852} {\bibfield  {journal} {\bibinfo  {journal}
  {J. Stat. Phys.}\ }\textbf {\bibinfo {volume} {98}},\ \bibinfo {pages} {973}
  (\bibinfo {year} {2000})},\ \Eprint
  {http://arxiv.org/abs/arXiv:cond-mat/9906076} {arXiv:cond-mat/9906076}
  \BibitemShut {NoStop}%
\bibitem [{\citenamefont {Joh}\ \emph {et~al.}(1999)\citenamefont {Joh},
  \citenamefont {Orbach}, \citenamefont {Wood}, \citenamefont {Hammann},\ and\
  \citenamefont {Vincent}}]{joh:99}%
  \BibitemOpen
  \bibfield  {author} {\bibinfo {author} {\bibfnamefont {Y.~G.}\ \bibnamefont
  {Joh}}, \bibinfo {author} {\bibfnamefont {R.}~\bibnamefont {Orbach}},
  \bibinfo {author} {\bibfnamefont {G.~G.}\ \bibnamefont {Wood}}, \bibinfo
  {author} {\bibfnamefont {J.}~\bibnamefont {Hammann}}, \ and\ \bibinfo
  {author} {\bibfnamefont {E.}~\bibnamefont {Vincent}},\ }\href {\doibase
  10.1103/PhysRevLett.82.438} {\bibfield  {journal} {\bibinfo  {journal} {Phys.
  Rev. Lett.}\ }\textbf {\bibinfo {volume} {82}},\ \bibinfo {pages} {438}
  (\bibinfo {year} {1999})}\BibitemShut {NoStop}%
\bibitem [{\citenamefont {Zhai}\ \emph {et~al.}(2017)\citenamefont {Zhai},
  \citenamefont {Harrison}, \citenamefont {Tennant}, \citenamefont {Dalhberg},
  \citenamefont {Kenning},\ and\ \citenamefont {Orbach}}]{zhai:17}%
  \BibitemOpen
  \bibfield  {author} {\bibinfo {author} {\bibfnamefont {Q.}~\bibnamefont
  {Zhai}}, \bibinfo {author} {\bibfnamefont {D.~C.}\ \bibnamefont {Harrison}},
  \bibinfo {author} {\bibfnamefont {D.}~\bibnamefont {Tennant}}, \bibinfo
  {author} {\bibfnamefont {E.~D.}\ \bibnamefont {Dalhberg}}, \bibinfo {author}
  {\bibfnamefont {G.~G.}\ \bibnamefont {Kenning}}, \ and\ \bibinfo {author}
  {\bibfnamefont {R.~L.}\ \bibnamefont {Orbach}},\ }\href {\doibase
  10.1103/PhysRevB.95.054304} {\bibfield  {journal} {\bibinfo  {journal} {Phys.
  Rev. B}\ }\textbf {\bibinfo {volume} {95}},\ \bibinfo {pages} {054304}
  (\bibinfo {year} {2017})}\BibitemShut {NoStop}%
\bibitem [{\citenamefont {Guchhait}\ and\ \citenamefont
  {Orbach}(2014)}]{guchhait:14}%
  \BibitemOpen
  \bibfield  {author} {\bibinfo {author} {\bibfnamefont {S.}~\bibnamefont
  {Guchhait}}\ and\ \bibinfo {author} {\bibfnamefont {R.}~\bibnamefont
  {Orbach}},\ }\href {\doibase 10.1103/PhysRevLett.112.126401} {\bibfield
  {journal} {\bibinfo  {journal} {Phys. Rev. Lett.}\ }\textbf {\bibinfo
  {volume} {112}},\ \bibinfo {pages} {126401} (\bibinfo {year}
  {2014})}\BibitemShut {NoStop}%
\bibitem [{\citenamefont {Guchhait}\ and\ \citenamefont
  {Orbach}(2017)}]{guchhait:17}%
  \BibitemOpen
  \bibfield  {author} {\bibinfo {author} {\bibfnamefont {S.}~\bibnamefont
  {Guchhait}}\ and\ \bibinfo {author} {\bibfnamefont {R.~L.}\ \bibnamefont
  {Orbach}},\ }\href@noop {} {\bibfield  {journal} {\bibinfo  {journal} {Phys.
  Rev. Lett.}\ }\textbf {\bibinfo {volume} {118}},\ \bibinfo {pages} {157203}
  (\bibinfo {year} {2017})}\BibitemShut {NoStop}%
\bibitem [{\citenamefont {Belletti}\ \emph {et~al.}(2008)\citenamefont
  {Belletti}, \citenamefont {Cotallo}, \citenamefont {Cruz}, \citenamefont
  {Fernandez}, \citenamefont {Gordillo-Guerrero}, \citenamefont {Guidetti},
  \citenamefont {Maiorano}, \citenamefont {Mantovani}, \citenamefont
  {Marinari}, \citenamefont {Mart\'{i}n-Mayor}, \citenamefont {Sudupe},
  \citenamefont {Navarro}, \citenamefont {Parisi}, \citenamefont
  {Perez-Gaviro}, \citenamefont {Ruiz-Lorenzo}, \citenamefont {Schifano},
  \citenamefont {Sciretti}, \citenamefont {Tarancon}, \citenamefont
  {Tripiccione}, \citenamefont {Velasco},\ and\ \citenamefont
  {Yllanes}}]{janus:08b}%
  \BibitemOpen
  \bibfield  {author} {\bibinfo {author} {\bibfnamefont {F.}~\bibnamefont
  {Belletti}}, \bibinfo {author} {\bibfnamefont {M.}~\bibnamefont {Cotallo}},
  \bibinfo {author} {\bibfnamefont {A.}~\bibnamefont {Cruz}}, \bibinfo {author}
  {\bibfnamefont {L.~A.}\ \bibnamefont {Fernandez}}, \bibinfo {author}
  {\bibfnamefont {A.}~\bibnamefont {Gordillo-Guerrero}}, \bibinfo {author}
  {\bibfnamefont {M.}~\bibnamefont {Guidetti}}, \bibinfo {author}
  {\bibfnamefont {A.}~\bibnamefont {Maiorano}}, \bibinfo {author}
  {\bibfnamefont {F.}~\bibnamefont {Mantovani}}, \bibinfo {author}
  {\bibfnamefont {E.}~\bibnamefont {Marinari}}, \bibinfo {author}
  {\bibfnamefont {V.}~\bibnamefont {Mart\'{i}n-Mayor}}, \bibinfo {author}
  {\bibfnamefont {A.~M.}\ \bibnamefont {Sudupe}}, \bibinfo {author}
  {\bibfnamefont {D.}~\bibnamefont {Navarro}}, \bibinfo {author} {\bibfnamefont
  {G.}~\bibnamefont {Parisi}}, \bibinfo {author} {\bibfnamefont
  {S.}~\bibnamefont {Perez-Gaviro}}, \bibinfo {author} {\bibfnamefont {J.~J.}\
  \bibnamefont {Ruiz-Lorenzo}}, \bibinfo {author} {\bibfnamefont {S.~F.}\
  \bibnamefont {Schifano}}, \bibinfo {author} {\bibfnamefont {D.}~\bibnamefont
  {Sciretti}}, \bibinfo {author} {\bibfnamefont {A.}~\bibnamefont {Tarancon}},
  \bibinfo {author} {\bibfnamefont {R.}~\bibnamefont {Tripiccione}}, \bibinfo
  {author} {\bibfnamefont {J.~L.}\ \bibnamefont {Velasco}}, \ and\ \bibinfo
  {author} {\bibfnamefont {D.}~\bibnamefont {Yllanes}} (\bibinfo
  {collaboration} {Janus Collaboration}),\ }\href {\doibase
  10.1103/PhysRevLett.101.157201} {\bibfield  {journal} {\bibinfo  {journal}
  {Phys. Rev. Lett.}\ }\textbf {\bibinfo {volume} {101}},\ \bibinfo {pages}
  {157201} (\bibinfo {year} {2008})},\ \Eprint
  {http://arxiv.org/abs/arXiv:0804.1471} {arXiv:0804.1471} \BibitemShut
  {NoStop}%
\bibitem [{\citenamefont {Lulli}\ \emph {et~al.}(2016)\citenamefont {Lulli},
  \citenamefont {Parisi},\ and\ \citenamefont {Pelissetto}}]{lulli:15}%
  \BibitemOpen
  \bibfield  {author} {\bibinfo {author} {\bibfnamefont {M.}~\bibnamefont
  {Lulli}}, \bibinfo {author} {\bibfnamefont {G.}~\bibnamefont {Parisi}}, \
  and\ \bibinfo {author} {\bibfnamefont {A.}~\bibnamefont {Pelissetto}},\
  }\href {\doibase 10.1103/PhysRevE.93.032126} {\bibfield  {journal} {\bibinfo
  {journal} {Phys. Rev. E}\ }\textbf {\bibinfo {volume} {93}},\ \bibinfo
  {pages} {032126} (\bibinfo {year} {2016})}\BibitemShut {NoStop}%
\bibitem [{\citenamefont {Belletti}\ \emph
  {et~al.}(2009{\natexlab{a}})\citenamefont {Belletti}, \citenamefont
  {Guidetti}, \citenamefont {Maiorano}, \citenamefont {Mantovani},
  \citenamefont {Schifano}, \citenamefont {Tripiccione}, \citenamefont
  {Cotallo}, \citenamefont {Perez-Gaviro}, \citenamefont {Sciretti},
  \citenamefont {Velasco}, \citenamefont {Cruz}, \citenamefont {Navarro},
  \citenamefont {Tarancon}, \citenamefont {Fernandez}, \citenamefont
  {Mart\'{i}n-Mayor}, \citenamefont {Mu{\~n}oz-Sudupe}, \citenamefont
  {Yllanes}, \citenamefont {Gordillo-Guerrero}, \citenamefont {Ruiz-Lorenzo},
  \citenamefont {Marinari}, \citenamefont {Parisi}, \citenamefont {Rossi},\
  and\ \citenamefont {Zanier}}]{janus:09}%
  \BibitemOpen
  \bibfield  {author} {\bibinfo {author} {\bibfnamefont {F.}~\bibnamefont
  {Belletti}}, \bibinfo {author} {\bibfnamefont {M.}~\bibnamefont {Guidetti}},
  \bibinfo {author} {\bibfnamefont {A.}~\bibnamefont {Maiorano}}, \bibinfo
  {author} {\bibfnamefont {F.}~\bibnamefont {Mantovani}}, \bibinfo {author}
  {\bibfnamefont {S.~F.}\ \bibnamefont {Schifano}}, \bibinfo {author}
  {\bibfnamefont {R.}~\bibnamefont {Tripiccione}}, \bibinfo {author}
  {\bibfnamefont {M.}~\bibnamefont {Cotallo}}, \bibinfo {author} {\bibfnamefont
  {S.}~\bibnamefont {Perez-Gaviro}}, \bibinfo {author} {\bibfnamefont
  {D.}~\bibnamefont {Sciretti}}, \bibinfo {author} {\bibfnamefont {J.~L.}\
  \bibnamefont {Velasco}}, \bibinfo {author} {\bibfnamefont {A.}~\bibnamefont
  {Cruz}}, \bibinfo {author} {\bibfnamefont {D.}~\bibnamefont {Navarro}},
  \bibinfo {author} {\bibfnamefont {A.}~\bibnamefont {Tarancon}}, \bibinfo
  {author} {\bibfnamefont {L.~A.}\ \bibnamefont {Fernandez}}, \bibinfo {author}
  {\bibfnamefont {V.}~\bibnamefont {Mart\'{i}n-Mayor}}, \bibinfo {author}
  {\bibfnamefont {A.}~\bibnamefont {Mu{\~n}oz-Sudupe}}, \bibinfo {author}
  {\bibfnamefont {D.}~\bibnamefont {Yllanes}}, \bibinfo {author} {\bibfnamefont
  {A.}~\bibnamefont {Gordillo-Guerrero}}, \bibinfo {author} {\bibfnamefont
  {J.~J.}\ \bibnamefont {Ruiz-Lorenzo}}, \bibinfo {author} {\bibfnamefont
  {E.}~\bibnamefont {Marinari}}, \bibinfo {author} {\bibfnamefont
  {G.}~\bibnamefont {Parisi}}, \bibinfo {author} {\bibfnamefont
  {M.}~\bibnamefont {Rossi}}, \ and\ \bibinfo {author} {\bibfnamefont
  {G.}~\bibnamefont {Zanier}} (\bibinfo {collaboration} {Janus
  Collaboration}),\ }\href {\doibase 10.1109/MCSE.2009.11} {\bibfield
  {journal} {\bibinfo  {journal} {Computing in Science and Engineering}\
  }\textbf {\bibinfo {volume} {11}},\ \bibinfo {pages} {48} (\bibinfo {year}
  {2009}{\natexlab{a}})}\BibitemShut {NoStop}%
\bibitem [{\citenamefont {Baity-Jesi}\ \emph {et~al.}(2012)\citenamefont
  {Baity-Jesi}, \citenamefont {Ba\~{n}os}, \citenamefont {Cruz}, \citenamefont
  {Fernandez}, \citenamefont {Gil-Narvion}, \citenamefont {Gordillo-Guerrero},
  \citenamefont {Guidetti}, \citenamefont {Iniguez}, \citenamefont {Maiorano},
  \citenamefont {Mantovani}, \citenamefont {Marinari}, \citenamefont
  {Mart\'{i}n-Mayor}, \citenamefont {Monforte-Garcia}, \citenamefont
  {Munoz~Sudupe}, \citenamefont {Navarro}, \citenamefont {Parisi},
  \citenamefont {Pivanti}, \citenamefont {Perez-Gaviro}, \citenamefont
  {Ricci-Tersenghi}, \citenamefont {Ruiz-Lorenzo}, \citenamefont {Schifano},
  \citenamefont {Seoane}, \citenamefont {Tarancon}, \citenamefont {Tellez},
  \citenamefont {Tripiccione},\ and\ \citenamefont {Yllanes}}]{janus:12b}%
  \BibitemOpen
  \bibfield  {author} {\bibinfo {author} {\bibfnamefont {M.}~\bibnamefont
  {Baity-Jesi}}, \bibinfo {author} {\bibfnamefont {R.~A.}\ \bibnamefont
  {Ba\~{n}os}}, \bibinfo {author} {\bibfnamefont {A.}~\bibnamefont {Cruz}},
  \bibinfo {author} {\bibfnamefont {L.~A.}\ \bibnamefont {Fernandez}}, \bibinfo
  {author} {\bibfnamefont {J.~M.}\ \bibnamefont {Gil-Narvion}}, \bibinfo
  {author} {\bibfnamefont {A.}~\bibnamefont {Gordillo-Guerrero}}, \bibinfo
  {author} {\bibfnamefont {M.}~\bibnamefont {Guidetti}}, \bibinfo {author}
  {\bibfnamefont {D.}~\bibnamefont {Iniguez}}, \bibinfo {author} {\bibfnamefont
  {A.}~\bibnamefont {Maiorano}}, \bibinfo {author} {\bibfnamefont
  {F.}~\bibnamefont {Mantovani}}, \bibinfo {author} {\bibfnamefont
  {E.}~\bibnamefont {Marinari}}, \bibinfo {author} {\bibfnamefont
  {V.}~\bibnamefont {Mart\'{i}n-Mayor}}, \bibinfo {author} {\bibfnamefont
  {J.}~\bibnamefont {Monforte-Garcia}}, \bibinfo {author} {\bibfnamefont
  {A.}~\bibnamefont {Munoz~Sudupe}}, \bibinfo {author} {\bibfnamefont
  {D.}~\bibnamefont {Navarro}}, \bibinfo {author} {\bibfnamefont
  {G.}~\bibnamefont {Parisi}}, \bibinfo {author} {\bibfnamefont
  {M.}~\bibnamefont {Pivanti}}, \bibinfo {author} {\bibfnamefont
  {S.}~\bibnamefont {Perez-Gaviro}}, \bibinfo {author} {\bibfnamefont
  {F.}~\bibnamefont {Ricci-Tersenghi}}, \bibinfo {author} {\bibfnamefont
  {J.~J.}\ \bibnamefont {Ruiz-Lorenzo}}, \bibinfo {author} {\bibfnamefont
  {S.~F.}\ \bibnamefont {Schifano}}, \bibinfo {author} {\bibfnamefont
  {B.}~\bibnamefont {Seoane}}, \bibinfo {author} {\bibfnamefont
  {A.}~\bibnamefont {Tarancon}}, \bibinfo {author} {\bibfnamefont
  {P.}~\bibnamefont {Tellez}}, \bibinfo {author} {\bibfnamefont
  {R.}~\bibnamefont {Tripiccione}}, \ and\ \bibinfo {author} {\bibfnamefont
  {D.}~\bibnamefont {Yllanes}},\ }\href {\doibase 10.1140/epjst/e2012-01636-9}
  {\bibfield  {journal} {\bibinfo  {journal} {Eur. Phys. J. Special Topics}\
  }\textbf {\bibinfo {volume} {{210}}},\ \bibinfo {pages} {{33}} (\bibinfo
  {year} {{2012}})},\ \Eprint {http://arxiv.org/abs/arXiv:1204.4134}
  {arXiv:1204.4134} \BibitemShut {NoStop}%
\bibitem [{\citenamefont {Baity-Jesi}\ \emph
  {et~al.}(2014{\natexlab{a}})\citenamefont {Baity-Jesi}, \citenamefont
  {Ba\~{n}os}, \citenamefont {Cruz}, \citenamefont {Fernandez}, \citenamefont
  {Gil-Narvion}, \citenamefont {Gordillo-Guerrero}, \citenamefont {Iniguez},
  \citenamefont {Maiorano}, \citenamefont {Mantovani}, \citenamefont
  {Marinari}, \citenamefont {Mart\'{i}n-Mayor}, \citenamefont
  {Monforte-Garcia}, \citenamefont {Mu{\~n}oz~Sudupe}, \citenamefont {Navarro},
  \citenamefont {Parisi}, \citenamefont {Perez-Gaviro}, \citenamefont
  {Pivanti}, \citenamefont {Ricci-Tersenghi}, \citenamefont {Ruiz-Lorenzo},
  \citenamefont {Schifano}, \citenamefont {Seoane}, \citenamefont {Tarancon},
  \citenamefont {Tripiccione},\ and\ \citenamefont {Yllanes}}]{janus:14}%
  \BibitemOpen
  \bibfield  {author} {\bibinfo {author} {\bibfnamefont {M.}~\bibnamefont
  {Baity-Jesi}}, \bibinfo {author} {\bibfnamefont {R.~A.}\ \bibnamefont
  {Ba\~{n}os}}, \bibinfo {author} {\bibfnamefont {A.}~\bibnamefont {Cruz}},
  \bibinfo {author} {\bibfnamefont {L.~A.}\ \bibnamefont {Fernandez}}, \bibinfo
  {author} {\bibfnamefont {J.~M.}\ \bibnamefont {Gil-Narvion}}, \bibinfo
  {author} {\bibfnamefont {A.}~\bibnamefont {Gordillo-Guerrero}}, \bibinfo
  {author} {\bibfnamefont {D.}~\bibnamefont {Iniguez}}, \bibinfo {author}
  {\bibfnamefont {A.}~\bibnamefont {Maiorano}}, \bibinfo {author}
  {\bibfnamefont {F.}~\bibnamefont {Mantovani}}, \bibinfo {author}
  {\bibfnamefont {E.}~\bibnamefont {Marinari}}, \bibinfo {author}
  {\bibfnamefont {V.}~\bibnamefont {Mart\'{i}n-Mayor}}, \bibinfo {author}
  {\bibfnamefont {J.}~\bibnamefont {Monforte-Garcia}}, \bibinfo {author}
  {\bibfnamefont {A.}~\bibnamefont {Mu{\~n}oz~Sudupe}}, \bibinfo {author}
  {\bibfnamefont {D.}~\bibnamefont {Navarro}}, \bibinfo {author} {\bibfnamefont
  {G.}~\bibnamefont {Parisi}}, \bibinfo {author} {\bibfnamefont
  {S.}~\bibnamefont {Perez-Gaviro}}, \bibinfo {author} {\bibfnamefont
  {M.}~\bibnamefont {Pivanti}}, \bibinfo {author} {\bibfnamefont
  {F.}~\bibnamefont {Ricci-Tersenghi}}, \bibinfo {author} {\bibfnamefont
  {J.~J.}\ \bibnamefont {Ruiz-Lorenzo}}, \bibinfo {author} {\bibfnamefont
  {S.~F.}\ \bibnamefont {Schifano}}, \bibinfo {author} {\bibfnamefont
  {B.}~\bibnamefont {Seoane}}, \bibinfo {author} {\bibfnamefont
  {A.}~\bibnamefont {Tarancon}}, \bibinfo {author} {\bibfnamefont
  {R.}~\bibnamefont {Tripiccione}}, \ and\ \bibinfo {author} {\bibfnamefont
  {D.}~\bibnamefont {Yllanes}} (\bibinfo {collaboration} {Janus
  Collaboration}),\ }\href {\doibase 10.1016/j.cpc.2013.10.019} {\bibfield
  {journal} {\bibinfo  {journal} {Comp. Phys. Comm}\ }\textbf {\bibinfo
  {volume} {185}},\ \bibinfo {pages} {550} (\bibinfo {year}
  {2014}{\natexlab{a}})},\ \Eprint {http://arxiv.org/abs/arXiv:1310.1032}
  {arXiv:1310.1032} \BibitemShut {NoStop}%
\bibitem [{\citenamefont {Baity-Jesi}\ \emph
  {et~al.}(2017{\natexlab{a}})\citenamefont {Baity-Jesi}, \citenamefont
  {Calore}, \citenamefont {Cruz}, \citenamefont {Fernandez}, \citenamefont
  {Gil-Narvi\'on}, \citenamefont {Gordillo-Guerrero}, \citenamefont {Iñiguez},
  \citenamefont {Maiorano}, \citenamefont {Marinari}, \citenamefont
  {Martin-Mayor}, \citenamefont {Monforte-Garcia}, \citenamefont
  {Muñoz~Sudupe}, \citenamefont {Navarro}, \citenamefont {Parisi},
  \citenamefont {Perez-Gaviro}, \citenamefont {Ricci-Tersenghi}, \citenamefont
  {Ruiz-Lorenzo}, \citenamefont {Schifano}, \citenamefont {Seoane},
  \citenamefont {Taranc\'on}, \citenamefont {Tripiccione},\ and\ \citenamefont
  {Yllanes}}]{janus:16}%
  \BibitemOpen
  \bibfield  {author} {\bibinfo {author} {\bibfnamefont {M.}~\bibnamefont
  {Baity-Jesi}}, \bibinfo {author} {\bibfnamefont {E.}~\bibnamefont {Calore}},
  \bibinfo {author} {\bibfnamefont {A.}~\bibnamefont {Cruz}}, \bibinfo {author}
  {\bibfnamefont {L.~A.}\ \bibnamefont {Fernandez}}, \bibinfo {author}
  {\bibfnamefont {J.~M.}\ \bibnamefont {Gil-Narvi\'on}}, \bibinfo {author}
  {\bibfnamefont {A.}~\bibnamefont {Gordillo-Guerrero}}, \bibinfo {author}
  {\bibfnamefont {D.}~\bibnamefont {Iñiguez}}, \bibinfo {author}
  {\bibfnamefont {A.}~\bibnamefont {Maiorano}}, \bibinfo {author}
  {\bibfnamefont {E.}~\bibnamefont {Marinari}}, \bibinfo {author}
  {\bibfnamefont {V.}~\bibnamefont {Martin-Mayor}}, \bibinfo {author}
  {\bibfnamefont {J.}~\bibnamefont {Monforte-Garcia}}, \bibinfo {author}
  {\bibfnamefont {A.}~\bibnamefont {Muñoz~Sudupe}}, \bibinfo {author}
  {\bibfnamefont {D.}~\bibnamefont {Navarro}}, \bibinfo {author} {\bibfnamefont
  {G.}~\bibnamefont {Parisi}}, \bibinfo {author} {\bibfnamefont
  {S.}~\bibnamefont {Perez-Gaviro}}, \bibinfo {author} {\bibfnamefont
  {F.}~\bibnamefont {Ricci-Tersenghi}}, \bibinfo {author} {\bibfnamefont
  {J.~J.}\ \bibnamefont {Ruiz-Lorenzo}}, \bibinfo {author} {\bibfnamefont
  {S.~F.}\ \bibnamefont {Schifano}}, \bibinfo {author} {\bibfnamefont
  {B.}~\bibnamefont {Seoane}}, \bibinfo {author} {\bibfnamefont
  {A.}~\bibnamefont {Taranc\'on}}, \bibinfo {author} {\bibfnamefont
  {R.}~\bibnamefont {Tripiccione}}, \ and\ \bibinfo {author} {\bibfnamefont
  {D.}~\bibnamefont {Yllanes}},\ }\href {\doibase 10.1073/pnas.1621242114}
  {\bibfield  {journal} {\bibinfo  {journal} {Proceedings of the National
  Academy of Sciences}\ }\textbf {\bibinfo {volume} {114}},\ \bibinfo {pages}
  {1838} (\bibinfo {year} {2017}{\natexlab{a}})}\BibitemShut {NoStop}%
\bibitem [{\citenamefont {Alvarez~Ba{\~n}os}\ \emph
  {et~al.}(2010{\natexlab{a}})\citenamefont {Alvarez~Ba{\~n}os}, \citenamefont
  {Cruz}, \citenamefont {Fernandez}, \citenamefont {Gil-Narvion}, \citenamefont
  {Gordillo-Guerrero}, \citenamefont {Guidetti}, \citenamefont {Maiorano},
  \citenamefont {Mantovani}, \citenamefont {Marinari}, \citenamefont
  {Mart\'{i}n-Mayor}, \citenamefont {Monforte-Garcia}, \citenamefont
  {Mu{\~n}oz~Sudupe}, \citenamefont {Navarro}, \citenamefont {Parisi},
  \citenamefont {Perez-Gaviro}, \citenamefont {Ruiz-Lorenzo}, \citenamefont
  {Schifano}, \citenamefont {Seoane}, \citenamefont {Tarancon}, \citenamefont
  {Tripiccione},\ and\ \citenamefont {Yllanes}}]{janus:10}%
  \BibitemOpen
  \bibfield  {author} {\bibinfo {author} {\bibfnamefont {R.}~\bibnamefont
  {Alvarez~Ba{\~n}os}}, \bibinfo {author} {\bibfnamefont {A.}~\bibnamefont
  {Cruz}}, \bibinfo {author} {\bibfnamefont {L.~A.}\ \bibnamefont {Fernandez}},
  \bibinfo {author} {\bibfnamefont {J.~M.}\ \bibnamefont {Gil-Narvion}},
  \bibinfo {author} {\bibfnamefont {A.}~\bibnamefont {Gordillo-Guerrero}},
  \bibinfo {author} {\bibfnamefont {M.}~\bibnamefont {Guidetti}}, \bibinfo
  {author} {\bibfnamefont {A.}~\bibnamefont {Maiorano}}, \bibinfo {author}
  {\bibfnamefont {F.}~\bibnamefont {Mantovani}}, \bibinfo {author}
  {\bibfnamefont {E.}~\bibnamefont {Marinari}}, \bibinfo {author}
  {\bibfnamefont {V.}~\bibnamefont {Mart\'{i}n-Mayor}}, \bibinfo {author}
  {\bibfnamefont {J.}~\bibnamefont {Monforte-Garcia}}, \bibinfo {author}
  {\bibfnamefont {A.}~\bibnamefont {Mu{\~n}oz~Sudupe}}, \bibinfo {author}
  {\bibfnamefont {D.}~\bibnamefont {Navarro}}, \bibinfo {author} {\bibfnamefont
  {G.}~\bibnamefont {Parisi}}, \bibinfo {author} {\bibfnamefont
  {S.}~\bibnamefont {Perez-Gaviro}}, \bibinfo {author} {\bibfnamefont {J.~J.}\
  \bibnamefont {Ruiz-Lorenzo}}, \bibinfo {author} {\bibfnamefont {S.~F.}\
  \bibnamefont {Schifano}}, \bibinfo {author} {\bibfnamefont {B.}~\bibnamefont
  {Seoane}}, \bibinfo {author} {\bibfnamefont {A.}~\bibnamefont {Tarancon}},
  \bibinfo {author} {\bibfnamefont {R.}~\bibnamefont {Tripiccione}}, \ and\
  \bibinfo {author} {\bibfnamefont {D.}~\bibnamefont {Yllanes}} (\bibinfo
  {collaboration} {Janus Collaboration}),\ }\href {\doibase
  10.1088/1742-5468/2010/06/P06026} {\bibfield  {journal} {\bibinfo  {journal}
  {J. Stat. Mech.}\ }\textbf {\bibinfo {volume} {2010}},\ \bibinfo {pages}
  {P06026} (\bibinfo {year} {2010}{\natexlab{a}})},\ \Eprint
  {http://arxiv.org/abs/arXiv:1003.2569} {arXiv:1003.2569} \BibitemShut
  {NoStop}%
\bibitem [{\citenamefont {Alvarez~Ba{\~n}os}\ \emph
  {et~al.}(2010{\natexlab{b}})\citenamefont {Alvarez~Ba{\~n}os}, \citenamefont
  {Cruz}, \citenamefont {Fernandez}, \citenamefont {Gil-Narvion}, \citenamefont
  {Gordillo-Guerrero}, \citenamefont {Guidetti}, \citenamefont {Maiorano},
  \citenamefont {Mantovani}, \citenamefont {Marinari}, \citenamefont
  {Mart\'{i}n-Mayor}, \citenamefont {Monforte-Garcia}, \citenamefont
  {Mu{\~n}oz~Sudupe}, \citenamefont {Navarro}, \citenamefont {Parisi},
  \citenamefont {Perez-Gaviro}, \citenamefont {Ruiz-Lorenzo}, \citenamefont
  {Schifano}, \citenamefont {Seoane}, \citenamefont {Tarancon}, \citenamefont
  {Tripiccione},\ and\ \citenamefont {Yllanes}}]{janus:10b}%
  \BibitemOpen
  \bibfield  {author} {\bibinfo {author} {\bibfnamefont {R.}~\bibnamefont
  {Alvarez~Ba{\~n}os}}, \bibinfo {author} {\bibfnamefont {A.}~\bibnamefont
  {Cruz}}, \bibinfo {author} {\bibfnamefont {L.~A.}\ \bibnamefont {Fernandez}},
  \bibinfo {author} {\bibfnamefont {J.~M.}\ \bibnamefont {Gil-Narvion}},
  \bibinfo {author} {\bibfnamefont {A.}~\bibnamefont {Gordillo-Guerrero}},
  \bibinfo {author} {\bibfnamefont {M.}~\bibnamefont {Guidetti}}, \bibinfo
  {author} {\bibfnamefont {A.}~\bibnamefont {Maiorano}}, \bibinfo {author}
  {\bibfnamefont {F.}~\bibnamefont {Mantovani}}, \bibinfo {author}
  {\bibfnamefont {E.}~\bibnamefont {Marinari}}, \bibinfo {author}
  {\bibfnamefont {V.}~\bibnamefont {Mart\'{i}n-Mayor}}, \bibinfo {author}
  {\bibfnamefont {J.}~\bibnamefont {Monforte-Garcia}}, \bibinfo {author}
  {\bibfnamefont {A.}~\bibnamefont {Mu{\~n}oz~Sudupe}}, \bibinfo {author}
  {\bibfnamefont {D.}~\bibnamefont {Navarro}}, \bibinfo {author} {\bibfnamefont
  {G.}~\bibnamefont {Parisi}}, \bibinfo {author} {\bibfnamefont
  {S.}~\bibnamefont {Perez-Gaviro}}, \bibinfo {author} {\bibfnamefont {J.~J.}\
  \bibnamefont {Ruiz-Lorenzo}}, \bibinfo {author} {\bibfnamefont {S.~F.}\
  \bibnamefont {Schifano}}, \bibinfo {author} {\bibfnamefont {B.}~\bibnamefont
  {Seoane}}, \bibinfo {author} {\bibfnamefont {A.}~\bibnamefont {Tarancon}},
  \bibinfo {author} {\bibfnamefont {R.}~\bibnamefont {Tripiccione}}, \ and\
  \bibinfo {author} {\bibfnamefont {D.}~\bibnamefont {Yllanes}} (\bibinfo
  {collaboration} {Janus Collaboration}),\ }\href {\doibase
  10.1103/PhysRevLett.105.177202} {\bibfield  {journal} {\bibinfo  {journal}
  {Phys. Rev. Lett.}\ }\textbf {\bibinfo {volume} {105}},\ \bibinfo {pages}
  {177202} (\bibinfo {year} {2010}{\natexlab{b}})},\ \Eprint
  {http://arxiv.org/abs/arXiv:1003.2943} {arXiv:1003.2943} \BibitemShut
  {NoStop}%
\bibitem [{\citenamefont {Baity-Jesi}\ \emph
  {et~al.}(2017{\natexlab{b}})\citenamefont {Baity-Jesi}, \citenamefont
  {Calore}, \citenamefont {Cruz}, \citenamefont {Fernandez}, \citenamefont
  {Gil-Narvion}, \citenamefont {Gordillo-Guerrero}, \citenamefont {I\~niguez},
  \citenamefont {Maiorano}, \citenamefont {Marinari}, \citenamefont
  {Martin-Mayor}, \citenamefont {Monforte-Garcia}, \citenamefont {Mu\~noz
  Sudupe}, \citenamefont {Navarro}, \citenamefont {Parisi}, \citenamefont
  {Perez-Gaviro}, \citenamefont {Ricci-Tersenghi}, \citenamefont
  {Ruiz-Lorenzo}, \citenamefont {Schifano}, \citenamefont {Seoane},
  \citenamefont {Tarancon}, \citenamefont {Tripiccione},\ and\ \citenamefont
  {Yllanes}}]{janus:17b}%
  \BibitemOpen
  \bibfield  {author} {\bibinfo {author} {\bibfnamefont {M.}~\bibnamefont
  {Baity-Jesi}}, \bibinfo {author} {\bibfnamefont {E.}~\bibnamefont {Calore}},
  \bibinfo {author} {\bibfnamefont {A.}~\bibnamefont {Cruz}}, \bibinfo {author}
  {\bibfnamefont {L.~A.}\ \bibnamefont {Fernandez}}, \bibinfo {author}
  {\bibfnamefont {J.~M.}\ \bibnamefont {Gil-Narvion}}, \bibinfo {author}
  {\bibfnamefont {A.}~\bibnamefont {Gordillo-Guerrero}}, \bibinfo {author}
  {\bibfnamefont {D.}~\bibnamefont {I\~niguez}}, \bibinfo {author}
  {\bibfnamefont {A.}~\bibnamefont {Maiorano}}, \bibinfo {author}
  {\bibfnamefont {E.}~\bibnamefont {Marinari}}, \bibinfo {author}
  {\bibfnamefont {V.}~\bibnamefont {Martin-Mayor}}, \bibinfo {author}
  {\bibfnamefont {J.}~\bibnamefont {Monforte-Garcia}}, \bibinfo {author}
  {\bibfnamefont {A.}~\bibnamefont {Mu\~noz Sudupe}}, \bibinfo {author}
  {\bibfnamefont {D.}~\bibnamefont {Navarro}}, \bibinfo {author} {\bibfnamefont
  {G.}~\bibnamefont {Parisi}}, \bibinfo {author} {\bibfnamefont
  {S.}~\bibnamefont {Perez-Gaviro}}, \bibinfo {author} {\bibfnamefont
  {F.}~\bibnamefont {Ricci-Tersenghi}}, \bibinfo {author} {\bibfnamefont
  {J.~J.}\ \bibnamefont {Ruiz-Lorenzo}}, \bibinfo {author} {\bibfnamefont
  {S.~F.}\ \bibnamefont {Schifano}}, \bibinfo {author} {\bibfnamefont
  {B.}~\bibnamefont {Seoane}}, \bibinfo {author} {\bibfnamefont
  {A.}~\bibnamefont {Tarancon}}, \bibinfo {author} {\bibfnamefont
  {R.}~\bibnamefont {Tripiccione}}, \ and\ \bibinfo {author} {\bibfnamefont
  {D.}~\bibnamefont {Yllanes}} (\bibinfo {collaboration} {Janus
  Collaboration}),\ }\href {\doibase 10.1103/PhysRevLett.118.157202} {\bibfield
   {journal} {\bibinfo  {journal} {Phys. Rev. Lett.}\ }\textbf {\bibinfo
  {volume} {118}},\ \bibinfo {pages} {157202} (\bibinfo {year}
  {2017}{\natexlab{b}})}\BibitemShut {NoStop}%
\bibitem [{\citenamefont {Wang}\ \emph {et~al.}(2017)\citenamefont {Wang},
  \citenamefont {Machta}, \citenamefont {Munoz-Bauza},\ and\ \citenamefont
  {Katzgraber}}]{wang:17}%
  \BibitemOpen
  \bibfield  {author} {\bibinfo {author} {\bibfnamefont {W.}~\bibnamefont
  {Wang}}, \bibinfo {author} {\bibfnamefont {J.}~\bibnamefont {Machta}},
  \bibinfo {author} {\bibfnamefont {H.}~\bibnamefont {Munoz-Bauza}}, \ and\
  \bibinfo {author} {\bibfnamefont {H.~G.}\ \bibnamefont {Katzgraber}},\ }\href
  {\doibase 10.1103/PhysRevB.96.184417} {\bibfield  {journal} {\bibinfo
  {journal} {Phys. Rev. B}\ }\textbf {\bibinfo {volume} {96}},\ \bibinfo
  {pages} {184417} (\bibinfo {year} {2017})}\BibitemShut {NoStop}%
\bibitem [{\citenamefont {Edwards}\ and\ \citenamefont
  {Anderson}(1975)}]{edwards:75}%
  \BibitemOpen
  \bibfield  {author} {\bibinfo {author} {\bibfnamefont {S.~F.}\ \bibnamefont
  {Edwards}}\ and\ \bibinfo {author} {\bibfnamefont {P.~W.}\ \bibnamefont
  {Anderson}},\ }\href {\doibase 10.1088/0305-4608/5/5/017} {\bibfield
  {journal} {\bibinfo  {journal} {Journal of Physics F: Metal Physics}\
  }\textbf {\bibinfo {volume} {5}},\ \bibinfo {pages} {965} (\bibinfo {year}
  {1975})}\BibitemShut {NoStop}%
\bibitem [{\citenamefont {Baity-Jesi}\ \emph {et~al.}(2013)\citenamefont
  {Baity-Jesi}, \citenamefont {Ba\~{n}os}, \citenamefont {Cruz}, \citenamefont
  {Fernandez}, \citenamefont {Gil-Narvion}, \citenamefont {Gordillo-Guerrero},
  \citenamefont {Iniguez}, \citenamefont {Maiorano}, \citenamefont {Mantovani},
  \citenamefont {Marinari}, \citenamefont {Mart\'{i}n-Mayor}, \citenamefont
  {Monforte-Garcia}, \citenamefont {Mu{\~n}oz~Sudupe}, \citenamefont {Navarro},
  \citenamefont {Parisi}, \citenamefont {Perez-Gaviro}, \citenamefont
  {Pivanti}, \citenamefont {Ricci-Tersenghi}, \citenamefont {Ruiz-Lorenzo},
  \citenamefont {Schifano}, \citenamefont {Seoane}, \citenamefont {Tarancon},
  \citenamefont {Tripiccione},\ and\ \citenamefont {Yllanes}}]{janus:13}%
  \BibitemOpen
  \bibfield  {author} {\bibinfo {author} {\bibfnamefont {M.}~\bibnamefont
  {Baity-Jesi}}, \bibinfo {author} {\bibfnamefont {R.~A.}\ \bibnamefont
  {Ba\~{n}os}}, \bibinfo {author} {\bibfnamefont {A.}~\bibnamefont {Cruz}},
  \bibinfo {author} {\bibfnamefont {L.~A.}\ \bibnamefont {Fernandez}}, \bibinfo
  {author} {\bibfnamefont {J.~M.}\ \bibnamefont {Gil-Narvion}}, \bibinfo
  {author} {\bibfnamefont {A.}~\bibnamefont {Gordillo-Guerrero}}, \bibinfo
  {author} {\bibfnamefont {D.}~\bibnamefont {Iniguez}}, \bibinfo {author}
  {\bibfnamefont {A.}~\bibnamefont {Maiorano}}, \bibinfo {author}
  {\bibfnamefont {F.}~\bibnamefont {Mantovani}}, \bibinfo {author}
  {\bibfnamefont {E.}~\bibnamefont {Marinari}}, \bibinfo {author}
  {\bibfnamefont {V.}~\bibnamefont {Mart\'{i}n-Mayor}}, \bibinfo {author}
  {\bibfnamefont {J.}~\bibnamefont {Monforte-Garcia}}, \bibinfo {author}
  {\bibfnamefont {A.}~\bibnamefont {Mu{\~n}oz~Sudupe}}, \bibinfo {author}
  {\bibfnamefont {D.}~\bibnamefont {Navarro}}, \bibinfo {author} {\bibfnamefont
  {G.}~\bibnamefont {Parisi}}, \bibinfo {author} {\bibfnamefont
  {S.}~\bibnamefont {Perez-Gaviro}}, \bibinfo {author} {\bibfnamefont
  {M.}~\bibnamefont {Pivanti}}, \bibinfo {author} {\bibfnamefont
  {F.}~\bibnamefont {Ricci-Tersenghi}}, \bibinfo {author} {\bibfnamefont
  {J.~J.}\ \bibnamefont {Ruiz-Lorenzo}}, \bibinfo {author} {\bibfnamefont
  {S.~F.}\ \bibnamefont {Schifano}}, \bibinfo {author} {\bibfnamefont
  {B.}~\bibnamefont {Seoane}}, \bibinfo {author} {\bibfnamefont
  {A.}~\bibnamefont {Tarancon}}, \bibinfo {author} {\bibfnamefont
  {R.}~\bibnamefont {Tripiccione}}, \ and\ \bibinfo {author} {\bibfnamefont
  {D.}~\bibnamefont {Yllanes}} (\bibinfo {collaboration} {Janus
  Collaboration}),\ }\href {\doibase 10.1103/PhysRevB.88.224416} {\bibfield
  {journal} {\bibinfo  {journal} {Phys. Rev. B}\ }\textbf {\bibinfo {volume}
  {88}},\ \bibinfo {pages} {224416} (\bibinfo {year} {{2013}})},\ \Eprint
  {http://arxiv.org/abs/arXiv:1310.2910} {arXiv:1310.2910} \BibitemShut
  {NoStop}%
\bibitem [{\citenamefont {Amit}\ and\ \citenamefont
  {Mart\'{i}n-Mayor}(2005)}]{amit:05}%
  \BibitemOpen
  \bibfield  {author} {\bibinfo {author} {\bibfnamefont {D.~J.}\ \bibnamefont
  {Amit}}\ and\ \bibinfo {author} {\bibfnamefont {V.}~\bibnamefont
  {Mart\'{i}n-Mayor}},\ }\href {\doibase 10.1142/9789812775313_bmatter} {\emph
  {\bibinfo {title} {Field Theory, the Renormalization Group and Critical
  Phenomena}}},\ \bibinfo {edition} {3rd}\ ed.\ (\bibinfo  {publisher} {World
  Scientific},\ \bibinfo {address} {Singapore},\ \bibinfo {year}
  {2005})\BibitemShut {NoStop}%
\bibitem [{\citenamefont {Yllanes}(2011)}]{yllanes:11}%
  \BibitemOpen
  \bibfield  {author} {\bibinfo {author} {\bibfnamefont {D.}~\bibnamefont
  {Yllanes}},\ }\emph {\bibinfo {title} {Rugged Free-Energy Landscapes in
  Disordered Spin Systems}},\ \href@noop {} {Ph.D. thesis},\ \bibinfo  {school}
  {Universidad Complutense de Madrid} (\bibinfo {year} {2011}),\ \Eprint
  {http://arxiv.org/abs/arXiv:1111.0266} {arXiv:1111.0266} \BibitemShut
  {NoStop}%
\bibitem [{\citenamefont {Belletti}\ \emph
  {et~al.}(2009{\natexlab{b}})\citenamefont {Belletti}, \citenamefont {Cruz},
  \citenamefont {Fernandez}, \citenamefont {Gordillo-Guerrero}, \citenamefont
  {Guidetti}, \citenamefont {Maiorano}, \citenamefont {Mantovani},
  \citenamefont {Marinari}, \citenamefont {Mart\'{i}n-Mayor}, \citenamefont
  {Monforte}, \citenamefont {Mu{\~n}oz~Sudupe}, \citenamefont {Navarro},
  \citenamefont {Parisi}, \citenamefont {Perez-Gaviro}, \citenamefont
  {Ruiz-Lorenzo}, \citenamefont {Schifano}, \citenamefont {Sciretti},
  \citenamefont {Tarancon}, \citenamefont {Tripiccione},\ and\ \citenamefont
  {Yllanes}}]{janus:09b}%
  \BibitemOpen
  \bibfield  {author} {\bibinfo {author} {\bibfnamefont {F.}~\bibnamefont
  {Belletti}}, \bibinfo {author} {\bibfnamefont {A.}~\bibnamefont {Cruz}},
  \bibinfo {author} {\bibfnamefont {L.~A.}\ \bibnamefont {Fernandez}}, \bibinfo
  {author} {\bibfnamefont {A.}~\bibnamefont {Gordillo-Guerrero}}, \bibinfo
  {author} {\bibfnamefont {M.}~\bibnamefont {Guidetti}}, \bibinfo {author}
  {\bibfnamefont {A.}~\bibnamefont {Maiorano}}, \bibinfo {author}
  {\bibfnamefont {F.}~\bibnamefont {Mantovani}}, \bibinfo {author}
  {\bibfnamefont {E.}~\bibnamefont {Marinari}}, \bibinfo {author}
  {\bibfnamefont {V.}~\bibnamefont {Mart\'{i}n-Mayor}}, \bibinfo {author}
  {\bibfnamefont {J.}~\bibnamefont {Monforte}}, \bibinfo {author}
  {\bibfnamefont {A.}~\bibnamefont {Mu{\~n}oz~Sudupe}}, \bibinfo {author}
  {\bibfnamefont {D.}~\bibnamefont {Navarro}}, \bibinfo {author} {\bibfnamefont
  {G.}~\bibnamefont {Parisi}}, \bibinfo {author} {\bibfnamefont
  {S.}~\bibnamefont {Perez-Gaviro}}, \bibinfo {author} {\bibfnamefont {J.~J.}\
  \bibnamefont {Ruiz-Lorenzo}}, \bibinfo {author} {\bibfnamefont {S.~F.}\
  \bibnamefont {Schifano}}, \bibinfo {author} {\bibfnamefont {D.}~\bibnamefont
  {Sciretti}}, \bibinfo {author} {\bibfnamefont {A.}~\bibnamefont {Tarancon}},
  \bibinfo {author} {\bibfnamefont {R.}~\bibnamefont {Tripiccione}}, \ and\
  \bibinfo {author} {\bibfnamefont {D.}~\bibnamefont {Yllanes}} (\bibinfo
  {collaboration} {Janus Collaboration}),\ }\href {\doibase
  10.1007/s10955-009-9727-z} {\bibfield  {journal} {\bibinfo  {journal} {J.
  Stat. Phys.}\ }\textbf {\bibinfo {volume} {135}},\ \bibinfo {pages} {1121}
  (\bibinfo {year} {2009}{\natexlab{b}})},\ \Eprint
  {http://arxiv.org/abs/arXiv:0811.2864} {arXiv:0811.2864} \BibitemShut
  {NoStop}%
\bibitem [{\citenamefont {{Janus Collaboration}}()}]{janus:xx}%
  \BibitemOpen
  \bibfield  {author} {\bibinfo {author} {\bibnamefont {{Janus
  Collaboration}}},\ }\href@noop {} {\bibinfo  {journal} {(in preparation)}\
  }\BibitemShut {NoStop}%
\bibitem [{\citenamefont {Berthier}\ \emph {et~al.}(2016)\citenamefont
  {Berthier}, \citenamefont {Charbonneau}, \citenamefont {Jin}, \citenamefont
  {Parisi}, \citenamefont {Seoane},\ and\ \citenamefont
  {Zamponi}}]{berthier:16}%
  \BibitemOpen
\bibfield  {journal} {  }\bibfield  {author} {\bibinfo {author} {\bibfnamefont
  {L.}~\bibnamefont {Berthier}}, \bibinfo {author} {\bibfnamefont
  {P.}~\bibnamefont {Charbonneau}}, \bibinfo {author} {\bibfnamefont
  {Y.}~\bibnamefont {Jin}}, \bibinfo {author} {\bibfnamefont {G.}~\bibnamefont
  {Parisi}}, \bibinfo {author} {\bibfnamefont {B.}~\bibnamefont {Seoane}}, \
  and\ \bibinfo {author} {\bibfnamefont {F.}~\bibnamefont {Zamponi}},\ }\href
  {\doibase 10.1073/pnas.1607730113} {\bibfield  {journal} {\bibinfo  {journal}
  {Proc. Natl. Acad. Sci. USA}\ }\textbf {\bibinfo {volume} {113}},\ \bibinfo
  {pages} {8397} (\bibinfo {year} {2016})}\BibitemShut {NoStop}%
\bibitem [{\citenamefont {Seoane}\ and\ \citenamefont
  {Zamponi}(ress)}]{seoane:18}%
  \BibitemOpen
  \bibfield  {author} {\bibinfo {author} {\bibfnamefont {B.}~\bibnamefont
  {Seoane}}\ and\ \bibinfo {author} {\bibfnamefont {F.}~\bibnamefont
  {Zamponi}},\ }\href {\doibase 10.1039/C8SM00859K} {\bibfield  {journal}
  {\bibinfo  {journal} {Soft Matter}\ } (\bibinfo {year} {2018, in press}),\
  10.1039/C8SM00859K}\BibitemShut {NoStop}%
\bibitem [{\citenamefont {Fern\'andez}\ and\ \citenamefont
  {Mart\'{i}n-Mayor}(2015)}]{fernandez:15}%
  \BibitemOpen
  \bibfield  {author} {\bibinfo {author} {\bibfnamefont {L.~A.}\ \bibnamefont
  {Fern\'andez}}\ and\ \bibinfo {author} {\bibfnamefont {V.}~\bibnamefont
  {Mart\'{i}n-Mayor}},\ }\href {\doibase 10.1103/PhysRevB.91.174202} {\bibfield
   {journal} {\bibinfo  {journal} {Phys. Rev. B}\ }\textbf {\bibinfo {volume}
  {91}},\ \bibinfo {pages} {174202} (\bibinfo {year} {2015})}\BibitemShut
  {NoStop}%
\bibitem [{\citenamefont {Fern\'andez}\ \emph {et~al.}(2018)\citenamefont
  {Fern\'andez}, \citenamefont {Marinari}, \citenamefont {Mart{\'i}n-Mayor},
  \citenamefont {Parisi},\ and\ \citenamefont {Ruiz-Lorenzo}}]{fernandez:18a}%
  \BibitemOpen
  \bibfield  {author} {\bibinfo {author} {\bibfnamefont {L.~A.}\ \bibnamefont
  {Fern\'andez}}, \bibinfo {author} {\bibfnamefont {E.}~\bibnamefont
  {Marinari}}, \bibinfo {author} {\bibfnamefont {V.}~\bibnamefont
  {Mart{\'i}n-Mayor}}, \bibinfo {author} {\bibfnamefont {G.}~\bibnamefont
  {Parisi}}, \ and\ \bibinfo {author} {\bibfnamefont {J.}~\bibnamefont
  {Ruiz-Lorenzo}},\ }\href@noop {} {\enquote {\bibinfo {title} {An
  experiment-oriented analysis of 2d spin-glass dynamics: a twelve time-decades
  scaling study},}\ } (\bibinfo {year} {2018}),\ \bibinfo {note} {submitted},\
  \Eprint {http://arxiv.org/abs/arXiv:1805.06738} {arXiv:1805.06738}
  \BibitemShut {NoStop}%
\bibitem [{Note1()}]{Note1}%
  \BibitemOpen
  \bibinfo {note} {Our result $z_\protect \text {c} =z(T\protect \tmspace
  -\thinmuskip {.1667em}=\protect \tmspace -\thinmuskip {.1667em}\protect
  \ensuremath {T_\protect \mathrm {c}}\protect \xspace ) = 6.69(6)$ is
  significantly more accurate that $z_\protect \text {c} = 6.80(15)$~\cite
  {lulli:15} and $z_\protect \text {c}=6.86(16)$~\cite {janus:08b}, even
  though, unlike Ref.~\cite {janus:08b}, we allow for corrections to scaling,
  which increases the statistical error, see Appendix~\ref
  {sec:parameters}}\BibitemShut {NoStop}%
\bibitem [{Note2()}]{Note2}%
  \BibitemOpen
  \bibinfo {note} {A naive explanation for the curvature in $\xi
  _{12}(T,\protect \ensuremath {t_\protect \mathrm {w}}\protect \xspace )$
  would be the existence of finite-size effects (see~\cite {janus:08b}).
  However, $c_2$ grows as we decrease $T$, while finite-size effects would be
  controlled by $\xi _{12}/L$, which is smaller for the lower temperatures. See
  Appendix~\ref {sec:finite_size_effects} for extensive checks that our $L=160$
  are safe.}\BibitemShut {Stop}%
\bibitem [{\citenamefont {McMillan}(1983)}]{mcmillan:83}%
  \BibitemOpen
  \bibfield  {author} {\bibinfo {author} {\bibfnamefont {W.~L.}\ \bibnamefont
  {McMillan}},\ }\href {\doibase 10.1103/PhysRevB.28.5216} {\bibfield
  {journal} {\bibinfo  {journal} {Phys. Rev. B}\ }\textbf {\bibinfo {volume}
  {28}},\ \bibinfo {pages} {5216} (\bibinfo {year} {1983})}\BibitemShut
  {NoStop}%
\bibitem [{\citenamefont {Bray}\ and\ \citenamefont {Moore}(1978)}]{bray:78}%
  \BibitemOpen
  \bibfield  {author} {\bibinfo {author} {\bibfnamefont {A.~J.}\ \bibnamefont
  {Bray}}\ and\ \bibinfo {author} {\bibfnamefont {M.~A.}\ \bibnamefont
  {Moore}},\ }\href {\doibase 10.1103/PhysRevLett.41.1068} {\bibfield
  {journal} {\bibinfo  {journal} {Phys. Rev. Lett.}\ }\textbf {\bibinfo
  {volume} {41}},\ \bibinfo {pages} {1068} (\bibinfo {year}
  {1978})}\BibitemShut {NoStop}%
\bibitem [{\citenamefont {Fisher}\ and\ \citenamefont
  {Huse}(1986)}]{fisher:86}%
  \BibitemOpen
  \bibfield  {author} {\bibinfo {author} {\bibfnamefont {D.~S.}\ \bibnamefont
  {Fisher}}\ and\ \bibinfo {author} {\bibfnamefont {D.~A.}\ \bibnamefont
  {Huse}},\ }\href {\doibase 10.1103/PhysRevLett.56.1601} {\bibfield  {journal}
  {\bibinfo  {journal} {Phys. Rev. Lett.}\ }\textbf {\bibinfo {volume} {56}},\
  \bibinfo {pages} {1601} (\bibinfo {year} {1986})}\BibitemShut {NoStop}%
\bibitem [{\citenamefont {Josephson}(1966)}]{josephson:66}%
  \BibitemOpen
  \bibfield  {author} {\bibinfo {author} {\bibfnamefont {B.~D.}\ \bibnamefont
  {Josephson}},\ }\href@noop {} {\bibfield  {journal} {\bibinfo  {journal}
  {Phys. Lett.}\ }\textbf {\bibinfo {volume} {21}},\ \bibinfo {pages} {608}
  (\bibinfo {year} {1966})}\BibitemShut {NoStop}%
\bibitem [{Note3()}]{Note3}%
  \BibitemOpen
  \bibinfo {note} {In theory, $\ell _\protect \text {J} \propto [1+j_0(\protect
  \ensuremath {T_\protect \mathrm {c}}\protect \xspace -T)^\nu +j_1(\protect
  \ensuremath {T_\protect \mathrm {c}}\protect \xspace -T)^{\omega \nu
  }](\protect \ensuremath {T_\protect \mathrm {c}}\protect \xspace -T)^{-\nu
  }$, where we include analytic ($j_0$) and confluent ($j_1)$ scaling
  corrections with $\omega =1.12(10)$~\cite {janus:13}. For Fig.~\ref
  {fig:replicon}, we have chosen $j_0$ and $j_1$ to obtain the best collapse
  for the lowest temperatures.}\BibitemShut {Stop}%
\bibitem [{\citenamefont {Boettcher}(2004)}]{boettcher:04}%
  \BibitemOpen
  \bibfield  {author} {\bibinfo {author} {\bibfnamefont {S.}~\bibnamefont
  {Boettcher}},\ }\href {\doibase 10.1140/epjb/e2004-00102-5} {\bibfield
  {journal} {\bibinfo  {journal} {Eur. Phys. J. B}\ }\textbf {\bibinfo {volume}
  {38}},\ \bibinfo {pages} {83} (\bibinfo {year} {2004})},\ \Eprint
  {http://arxiv.org/abs/arXiv:cond-mat/0310698} {arXiv:cond-mat/0310698}
  \BibitemShut {NoStop}%
\bibitem [{\citenamefont {Bouchaud}\ \emph {et~al.}(2001)\citenamefont
  {Bouchaud}, \citenamefont {Dupuis}, \citenamefont {Hammann},\ and\
  \citenamefont {Vincent}}]{bouchaud:01}%
  \BibitemOpen
  \bibfield  {author} {\bibinfo {author} {\bibfnamefont {J.-P.}\ \bibnamefont
  {Bouchaud}}, \bibinfo {author} {\bibfnamefont {V.}~\bibnamefont {Dupuis}},
  \bibinfo {author} {\bibfnamefont {J.}~\bibnamefont {Hammann}}, \ and\
  \bibinfo {author} {\bibfnamefont {E.}~\bibnamefont {Vincent}},\ }\href
  {\doibase 10.1103/PhysRevB.65.024439} {\bibfield  {journal} {\bibinfo
  {journal} {Phys. Rev. B}\ }\textbf {\bibinfo {volume} {65}},\ \bibinfo
  {pages} {024439} (\bibinfo {year} {2001})}\BibitemShut {NoStop}%
\bibitem [{\citenamefont {Berthier}\ and\ \citenamefont
  {Bouchaud}(2002)}]{berthier:02}%
  \BibitemOpen
  \bibfield  {author} {\bibinfo {author} {\bibfnamefont {L.}~\bibnamefont
  {Berthier}}\ and\ \bibinfo {author} {\bibfnamefont {J.-P.}\ \bibnamefont
  {Bouchaud}},\ }\href {\doibase 10.1103/PhysRevB.66.054404} {\bibfield
  {journal} {\bibinfo  {journal} {Phys. Rev. B}\ }\textbf {\bibinfo {volume}
  {66}},\ \bibinfo {pages} {054404} (\bibinfo {year} {2002})}\BibitemShut
  {NoStop}%
\bibitem [{\citenamefont {Schins}\ \emph {et~al.}(1993)\citenamefont {Schins},
  \citenamefont {Arts},\ and\ \citenamefont {de~Wijn}}]{schins:93}%
  \BibitemOpen
  \bibfield  {author} {\bibinfo {author} {\bibfnamefont {A.~G.}\ \bibnamefont
  {Schins}}, \bibinfo {author} {\bibfnamefont {A.~F.~M.}\ \bibnamefont {Arts}},
  \ and\ \bibinfo {author} {\bibfnamefont {H.~W.}\ \bibnamefont {de~Wijn}},\
  }\href {\doibase 10.1103/PhysRevLett.70.2340} {\bibfield  {journal} {\bibinfo
   {journal} {Phys. Rev. Lett.}\ }\textbf {\bibinfo {volume} {70}},\ \bibinfo
  {pages} {2340} (\bibinfo {year} {1993})}\BibitemShut {NoStop}%
\bibitem [{\citenamefont {Rieger}(1993)}]{rieger:93}%
  \BibitemOpen
  \bibfield  {author} {\bibinfo {author} {\bibfnamefont {H.}~\bibnamefont
  {Rieger}},\ }\href {\doibase 10.1088/0305-4470/26/15/001} {\bibfield
  {journal} {\bibinfo  {journal} {J. Phys. A}\ }\textbf {\bibinfo {volume}
  {26}},\ \bibinfo {pages} {L615} (\bibinfo {year} {1993})}\BibitemShut
  {NoStop}%
\bibitem [{Note4()}]{Note4}%
  \BibitemOpen
  \bibinfo {note} {$G(T)$ in~\protect \textup {\hbox {\mathsurround \z@
  \protect \normalfont (\ignorespaces \ref {eq:divergente}\unskip \@@italiccorr
  )}} goes to zero at \protect \ensuremath {T_\protect \mathrm {c}}\protect
  \xspace as $(\protect \ensuremath {T_\protect \mathrm {c}}\protect \xspace
  -T)^{\Psi \nu }$, which is another form of the Josephson
  scaling.}\BibitemShut {Stop}%
\bibitem [{\citenamefont {Billoire}\ and\ \citenamefont
  {Marinari}(2001)}]{billoire:01}%
  \BibitemOpen
  \bibfield  {author} {\bibinfo {author} {\bibfnamefont {A.}~\bibnamefont
  {Billoire}}\ and\ \bibinfo {author} {\bibfnamefont {E.}~\bibnamefont
  {Marinari}},\ }\href {\doibase 10.1088/0305-4470/34/50/101} {\bibfield
  {journal} {\bibinfo  {journal} {Journal of Physics A: Mathematical and
  General}\ }\textbf {\bibinfo {volume} {34}},\ \bibinfo {pages} {L727}
  (\bibinfo {year} {2001})}\BibitemShut {NoStop}%
\bibitem [{\citenamefont {Rogers}\ and\ \citenamefont
  {Moore}(1989)}]{rodgers:89}%
  \BibitemOpen
  \bibfield  {author} {\bibinfo {author} {\bibfnamefont {G.}~\bibnamefont
  {Rogers}}\ and\ \bibinfo {author} {\bibfnamefont {M.~A.}\ \bibnamefont
  {Moore}},\ }\href {\doibase 10.1088/0305-4470/22/8/022} {\bibfield  {journal}
  {\bibinfo  {journal} {J. Phys. A: Math. Gen.}\ }\textbf {\bibinfo {volume}
  {22}},\ \bibinfo {pages} {1085} (\bibinfo {year} {1989})}\BibitemShut
  {NoStop}%
\bibitem [{\citenamefont {Colborne}(1990)}]{colborne:90}%
  \BibitemOpen
  \bibfield  {author} {\bibinfo {author} {\bibfnamefont {S.~G.~W.}\
  \bibnamefont {Colborne}},\ }\href {\doibase 10.1088/0305-4470/23/17/030}
  {\bibfield  {journal} {\bibinfo  {journal} {J. Phys. A: Math. Gen.}\ }\textbf
  {\bibinfo {volume} {23}},\ \bibinfo {pages} {4013} (\bibinfo {year}
  {1990})}\BibitemShut {NoStop}%
\bibitem [{\citenamefont {Bray}\ and\ \citenamefont {Moore}(1982)}]{bray:82}%
  \BibitemOpen
  \bibfield  {author} {\bibinfo {author} {\bibfnamefont {A.~J.}\ \bibnamefont
  {Bray}}\ and\ \bibinfo {author} {\bibfnamefont {M.~A.}\ \bibnamefont
  {Moore}},\ }\href {\doibase 10.1088/0022-3719/15/18/007} {\bibfield
  {journal} {\bibinfo  {journal} {J. Phys. C: Solid St. Phys.}\ }\textbf
  {\bibinfo {volume} {15}},\ \bibinfo {pages} {3897} (\bibinfo {year}
  {1982})}\BibitemShut {NoStop}%
\bibitem [{\citenamefont {Baity-Jesi}\ \emph
  {et~al.}(2014{\natexlab{b}})\citenamefont {Baity-Jesi}, \citenamefont
  {Fernandez}, \citenamefont {Mart\'{i}n-Mayor},\ and\ \citenamefont
  {Sanz}}]{baityjesi:14}%
  \BibitemOpen
  \bibfield  {author} {\bibinfo {author} {\bibfnamefont {M.}~\bibnamefont
  {Baity-Jesi}}, \bibinfo {author} {\bibfnamefont {L.~A.}\ \bibnamefont
  {Fernandez}}, \bibinfo {author} {\bibfnamefont {V.}~\bibnamefont
  {Mart\'{i}n-Mayor}}, \ and\ \bibinfo {author} {\bibfnamefont {J.~M.}\
  \bibnamefont {Sanz}},\ }\href {\doibase 10.1103/PhysRevB.89.014202}
  {\bibfield  {journal} {\bibinfo  {journal} {Phys. Rev.}\ }\textbf {\bibinfo
  {volume} {89}},\ \bibinfo {pages} {014202} (\bibinfo {year}
  {2014}{\natexlab{b}})},\ \Eprint {http://arxiv.org/abs/arXiv:1309.1599}
  {arXiv:1309.1599} \BibitemShut {NoStop}%
\end{thebibliography}
\end{document}